%% file: main.tex
\documentclass[conference,10pt]{IEEEtran}

\newif{\ifshowcomments}
\newif{\ifshowauthors}

\showauthorstrue

\input{preamble}

\title{MLIR for Quantum Beyond Gate Cancellation:\\Quantum Circuit Mapping Reimagined}

\ifshowauthors
\author{
    \IEEEauthorblockN{
        Matthias Reumann\IEEEauthorrefmark{1},
        Yannick Stade\IEEEauthorrefmark{1},
        Robert Wille\IEEEauthorrefmark{1}\IEEEauthorrefmark{2}, and
        Lukas Burgholzer\IEEEauthorrefmark{1}\IEEEauthorrefmark{2}
    }
    \IEEEauthorblockA{\IEEEauthorrefmark{1}%
    Chair for Design Automation,
        Technical University of Munich,
        Munich, Germany
    }
    \IEEEauthorblockA{\IEEEauthorrefmark{2}%
    Munich Quantum Software Company GmbH,
        Garching near Munich, Germany
    }
    \{%
    matthias.reumann, %
    yannick.stade, %
    robert.wille, %
    lukas.burgholzer\}@tum.de\\
    \href{https://www.cda.cit.tum.de/research/quantum}{www.cda.cit.tum.de/research/quantum}
}
\hypersetup{
    pdftitle={MLIR for Quantum Beyond Gate Cancellation: Quantum Circuit Mapping Reimagined},
    pdfsubject={},
    pdfauthor={
        Matthias Reumann, %
        Yannick Stade, %
        Robert Wille, and %
        Lukas Burgholzer
    }
}
\else
\author{%
    \vspace{6em}
}
\hypersetup{ 
    pdftitle={},
    pdfsubject={},
    pdfauthor={}
}
\fi

\begin{document}

    \maketitle

    \begin{abstract}
        The Multi-Level Intermediate Representation (MLIR) framework has become a cornerstone for building extensible, domain-specific compilers, with the quantum computing community already leveraging it to model quantum programs and implement basic optimizations. However, computationally intensive tasks in the quantum compilation pipeline, such as quantum circuit mapping, remain underexplored within the MLIR ecosystem. This paper proposes an MLIR-native blueprint for these non-local, quantum-specific optimization routines by reimplementing a well-established, state-of-the-art mapping A* search algorithm for qubit routing and SWAP insertion. Our evaluation demonstrates that this approach not only integrates seamlessly into an MLIR-based quantum compiler collection but also surpasses previous non-MLIR solutions in both solution quality and runtime. The implementation is open-source and publicly available at~\href{https://github.com/munich-quantum-toolkit/core}{github.com/munich-quantum-toolkit/core}.
    \end{abstract}

    \begin{IEEEkeywords}
    quantum computing, quantum circuit mapping, quantum compilation
    \end{IEEEkeywords}

    \section{Introduction}\label{sec:introduction}

    As quantum computing hardware scales beyond the noisy intermediate-scale quantum (NISQ) era~\cite{preskillQuantumComputingNISQ2018, preskill2025megaquop}, the complexity of quantum programs that can be executed increases significantly.
    However, the inherent noise and limited coherence times of current and near-term devices necessitate highly optimized compilation pipelines to maximize circuit fidelity.
    While simple, local optimizations such as gate cancellation or fusion are standard in most compilers, they are increasingly insufficient to bridge the gap between algorithmic requirements and hardware capabilities.
    Consequently, there is a growing need for advanced, global optimization routines capable of significantly reducing circuit depth and error rates.

    Developing such sophisticated compilers from scratch is a formidable software engineering challenge.
    Instead of reinventing the wheel, the classical compiler community has converged on Multi-Level Intermediate Representation (MLIR)~\cite{mlir} as a modular and extensible framework for building domain-specific compilers.
    Following its immense success in the machine learning and high-performance computing domains~\cite{triton2019, Abadi_TensorFlow_Large-scale_machine_2015, paszke2019pytorchimperativestylehighperformance, jin2020compilingonnxneuralnetwork}, MLIR is gaining traction in the quantum computing community.
    Several works have explored modeling quantum programs within MLIR and implementing basic optimizations~\cite{ittah_qiro_2022, peduri_qssa_2022, mccaskey_mlir_2021, ittah_catalyst_2024, healy_design_2024}.
    However, these efforts largely focus on feasibility studies involving relatively simple, local transformations.
    A crucial question remains unanswered: Is MLIR suitable for implementing complex, quantum-specific optimization routines that are non-local and computationally intensive?

    In this work, we address this question by investigating quantum circuit mapping ---often synonymously referred to as qubit routing or layout synthesis in the literature---within the MLIR ecosystem.
    Quantum circuit mapping is a mandatory and critical step in the compilation pipeline for architectures with restricted qubit connectivity, such as superconducting quantum processors.
    The problem involves transforming an abstract quantum circuit into one that respects the hardware's coupling graph by inserting SWAP gates, aiming to minimize the added overhead.
    Since the problem is NP-complete and the search space grows exponentially with the number of qubits, it represents a significantly more challenging test case for MLIR than typical peephole optimizations.

    With this work, we do not aim to propose yet another mapping algorithm, as the literature is already rich with solutions.
    Instead, we implement a well-established, state-of-the-art mapping approach based on A* search~\cite{zulehnerEfficientMethodologyMapping2019} using MLIR-native abstractions.
    This allows us to critically evaluate whether the framework's infrastructure facilitates the development of such complex algorithms and how the resulting performance compares to bespoke C++ implementations.

    Our evaluation demonstrates that:
    \begin{itemize}
        \item The investigated mapping algorithm integrates seamlessly into an MLIR-native compiler collection, confirming that MLIR provides the necessary tools and abstractions for complex quantum optimizations.
        \item Leveraging existing MLIR infrastructure and utilities significantly reduces development time, allowing developers to focus on algorithms rather than boilerplate code.
        \item The MLIR-native implementation outperforms the original state-of-the-art C++ implementation in terms of both solution quality (fewer SWAP gates) and runtime.
    \end{itemize}
    Ultimately, this work provides evidence that MLIR is not just a viable modeling tool but a powerful and efficient framework for the next generation of high-performance quantum compilers.

    This paper is structured as follows: 
    To ensure that the paper is self-contained, we review the necessary background in Section~\ref{sec:background}. 
    Section~\ref{sec:motivation} contextualizes and motivates this work within the existing literature. 
    In Section~\ref{sec:proposed-solution}, we outline the proposed solution in detail and subsequently evaluate it in Section~\ref{sec:evaluation}. 
    Finally, Section~\ref{sec:conclusion} concludes the paper.

    \clearpage

    \section{Background}\label{sec:background}
    This section outlines the fundamental concepts required to understand the remainder of this paper.

    \subsection{Multi-Level Intermediate Representation (MLIR)}\label{sec:mlir}

    Building an end-to-end quantum compilation pipeline entails substantial engineering effort.
    The underlying infrastructure must efficiently model, parse, and transform quantum programs while addressing both current and emerging requirements.

    Whereas previous efforts often involved building quantum compilers from scratch~\cite{Sivarajah_TKET_A_Retargetable_2020, qiskit2024}, a recent trend has shifted toward a \enquote{classical-first} approach~\cite{burgholzer2026mqtcompilercollectionblueprint}. 
    This strategy leverages established classical compiler infrastructures, such as the Multi-Level Intermediate Representation (MLIR)~\cite{mlir}, to support quantum computing. 
    This section provides an overview of modeling and transforming quantum programs within the MLIR framework.

    The MQT Compiler Collection~\cite{burgholzer2026mqtcompilercollectionblueprint} models quantum programs as directed acyclic \enquote{data-flow} graphs. 
    Each quantum operation is represented as a vertex that consumes and produces qubit values (represented as ingoing and outgoing edges, respectively). 
    Hence, each edge captures the evolution of a qubit's state throughout the circuit.
    This representation is particularly advantageous for optimization and complex transformations, as it makes dependencies between operations explicit.\matthias{Basierend auf der proposed solution, bei der der data-flow graph basically die wichtigste rolle dreh ich hier mal den spieß um und start mit der graph representation anstelle von "dialect is...". Bin gespannt was ihr sagt.}

    Moreover, each qubit state is described as a Static Single Assignment (SSA) value. 
    By definition, an SSA value is assigned exactly once, ensuring that the data flow remains immutable and explicitly traceable. 
    Within the MLIR framework, these values are denoted by the \texttt{\%} prefix.

    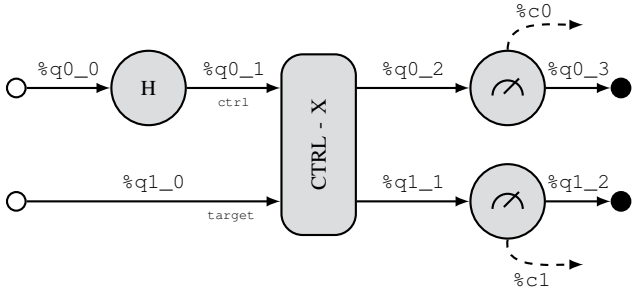
\begin{figure}[h!]
        \centering
        \input{figures/dataflow}
        \caption{The data-flow graph of a two qubit circuit.}
        \label{fig:dataflow}
    \end{figure}
    
    \begin{example}
       \autoref{fig:dataflow} illustrates the data-flow graph of a two-qubit circuit. 
       As shown, allocations ($\circ$) initialize the circuit by producing the initial SSA values \texttt{\%q0\_0} and \texttt{\%q1\_0}. 
       A subsequent Hadamard transforms the first qubit, consuming \texttt{\%q0\_0} and producing \texttt{\%q0\_1}. 
       Similarly, a controlled-X consumes \texttt{\%q0\_1} and \texttt{\%q1\_0} to produce \texttt{\%q0\_2} and \texttt{\%q1\_1}. 
       Measurements follow a similar logic, with the addition that they also produce classical SSA values, \texttt{\%c0} and \texttt{\%c1}. 
       Finally, sinks ($\bullet$) consume the final qubit values, thereby defining the end of their respective lifetimes.
    \end{example}

    To implement this domain-specific language, the MQT Compiler Collection leverages MLIR’s dialect mechanism. 
    A dialect groups a set of operations, types, and attributes under a common namespace. 
    While MLIR provides various built-in dialects, its architecture is fundamentally extensible, offering a framework for developing custom dialects tailored to specialized domains. 
    Furthermore, MLIR natively supports the parsing and transformation of IR modules that integrate primitives from multiple coexisting dialects.
    The Quantum Circuit Optimization (QCO) dialect, introduced within the MQT Compiler Collection, defines the primitive types (e.g., \texttt{!qco.qubit}) and operations (e.g., \texttt{qco.h}) necessary for modeling quantum computations. 

    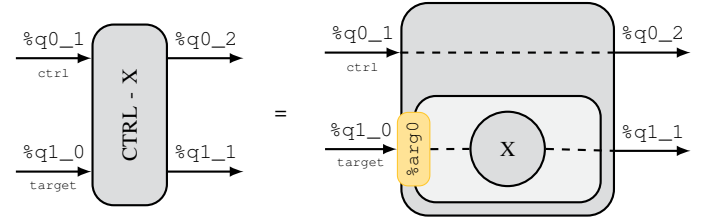
\begin{figure}[h!]
        \centering
        \input{figures/modifier}
        \caption{An illustration of the control-modifier in the QCO dialect.}
        \label{fig:modifier}
    \end{figure}
    
    \begin{example}
        The following snippet presents the textual intermediate representation (IR) corresponding to the data-flow graph in \autoref{fig:dataflow}. 
        The controlled-X gate is realized via the \texttt{qco.ctrl} modifier, a generic construct that enables the definition of arbitrary controlled operations. 
        The \texttt{qco.yield} operation serves as the terminator for the nested region (delimited by curly brackets), passing the transformed qubit values back to the outer scope. 
        \autoref{fig:modifier} provides a visual schematic of the data-flow within this control-modifier construct.
    \end{example}

    \begin{Verbatim}[fontsize=\footnotesize, frame=lines, framesep=2mm, baselinestretch=1.2]
%q0_0 = qco.alloc : !qco.qubit
%q1_0 = qco.alloc : !qco.qubit

%q0_1 = qco.h %q0_0 : !qco.qubit -> !qco.qubit
%q0_2, %q1_1 = qco.ctrl(%q0_1) 
                   targets (%arg0 = %q1_0)
{
    %q1_1 = qco.x %arg0 : !qco.qubit -> !qco.qubit
    qco.yield %q1_1
} : ({!qco.qubit}, {!qco.qubit}) ->
    ({!qco.qubit}, {!qco.qubit})

%q0_3, %c0 = qco.measure %q0_2 : !qco.qubit
%q1_2, %c1 = qco.measure %q1_1 : !qco.qubit

qco.sink %q0_3 : !qco.qubit
qco.sink %q1_2 : !qco.qubit
    \end{Verbatim}

    MLIR provides two primary methods for traversing IRs:
    \begin{itemize} 
    \item \textbf{Structural perspective}: This involves walking the IR operations in hierarchical order (top-down or bottom-up) and recursively visiting nested regions, such as those within \texttt{qco.ctrl}. 
    MLIR provides utilities to perform these traversals efficiently. 
    
    \item \textbf{Data-flow perspective}: This involves traversing the data-flow graph (as illustrated in \autoref{fig:dataflow}) by following MLIR's def-use chains. 
    These chains link an SSA value to the operation that defined it and the subsequent operations that consume it (the \enquote{users}). 
    While MLIR supports standard SSA data-flow, it lacks built-in support for traversing these chains according to specific quantum semantics.    
    \end{itemize}

    \begin{example}
        For instance, in the circuit shown in \autoref{fig:dataflow}, the defining operation for the SSA value \texttt{\%q0\_1} is the Hadamard gate, while its consumer (or user) is the controlled-X operation.
    \end{example}

    The MLIR framework provides comprehensive utilities for manipulating IRs by inserting, updating, or deleting operations. These transformations are orchestrated by its pass infrastructure, which provides a generic mechanism for analysis and rewriting. In this context, a pass traverses the IR to perform analysis or transformation. Furthermore, individual passes can be composed into a pass pipeline, ensuring they are executed in a deterministic sequence. Much like its dialect system, MLIR’s pass infrastructure is highly extensible, allowing for the development of custom transformations tailored to specific quantum compilation requirements.

    \begin{example}
    The canonicalization pass of the QCO dialect implements a custom rewrite strategy that removes unused qubits. To find these, MLIR's def-use chain is utilized: If the defining operation of the input of the \texttt{qco.sink} operation is a \texttt{qco.alloc} operation, both operations can be removed.
    \end{example}

    In summary, the MLIR framework and the MQT Compiler Collection's QCO dialect provide the necessary foundation to implement an efficient quantum circuit mapping pass.

    \subsection{Quantum Circuit Mapping using A* search}\label{sec:mapping}

    Each quantum computing architecture imposes a unique set of hardware-specific constraints. 
    Consequently, the compiler must transform high-level quantum programs into a format that adheres to these constraints while maintaining logical equivalence.

    In superconducting architectures, for instance, two-qubit connectivity is typically restricted: gates can only be executed between physically adjacent qubits. 
    Consequently, a necessary precondition for mapping is the decomposition in one and two-qubit gates (excluding barriers).
    These limitations are formally represented by a coupling graph $G=(V, E)$, where vertices $V$ denote hardware qubits and edges $E$ represent the available two-qubit interactions.

    \begin{figure}[h!]
        \centering
        \input{figures/novera}
        \caption{A $3\times3$ grid architecture.}
        \label{fig:novera}
    \end{figure}
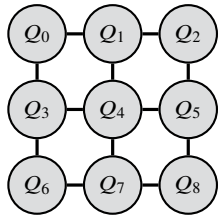
    
    \begin{example}
        \autoref{fig:novera} depicts the coupling graph for a superconducting processor, where physical qubits are arranged in a $3\times3$ grid.
        Two-qubit operations are restricted to adjacent qubits that are connected with an edge. 
        For example, while a two-qubit gate can be directly applied to hardware qubits $Q_0$ and $Q_3$, a direct interaction between $Q_0$ and $Q_4$ is prohibited by the architecture.
    \end{example}

    Due to these topological constraints, many quantum programs cannot be executed natively on a given target architecture.
    This occurs whenever no perfect initial program-to-hardware qubit mapping exists, for which the target qubits of all two-qubit gates are physically adjacent on the hardware coupling graph.

   A naive solution to this problem involves inserting SWAP gates along the shortest path between non-adjacent target qubits. 
   This process progressively moves the state of one target qubit until it is physically adjacent to the other, enabling the execution of the required gate.

    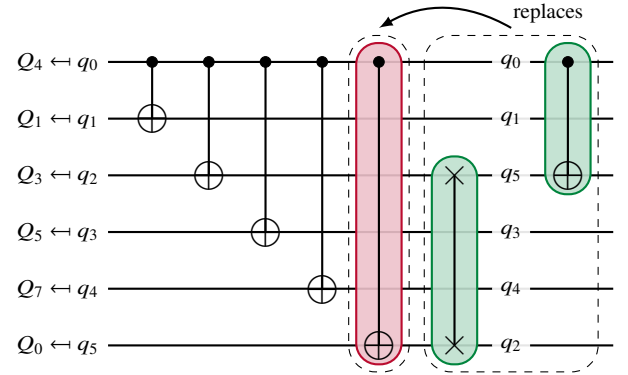
\begin{figure}[h!]
        \centering
        \input{figures/mapping}
        \caption{By exchanging the program qubits $q_5$ and $q_2$ with a SWAP gate, the final controlled-X gate acts on the adjacent hardware qubits $Q_4$ and $Q_3$.}
        \label{fig:mapping}
    \end{figure}

    \begin{example}
        \autoref{fig:mapping} illustrates the mapping procedure for a 6-qubit circuit onto the architecture from~\autoref{fig:novera}.
        Initially, the program qubits $(q_0, q_1, q_2, q_3, q_4, q_5)$ are assigned to the physical qubits $(Q_4, Q_1, Q_3, Q_5, Q_7, Q_0)$, respectively.
        Because program qubit $q_0$ has an output degree of $5$ and the maximum output degree of the architecture is $4$ ($Q_4$), no static mapping can satisfy all connectivity constraints. 
        Consequently, certain two-qubit gates remain non-executable.
        For example, the controlled-X gate (highlighted in red) initially acts on a non-adjacent pair of hardware qubits. 
        To resolve this, a SWAP operation is performed between the hardware qubits $Q_0$ and $Q_3$, exchanging the locations of the program qubits $q_2$ and $q_5$. 
        Subsequently, the final gate is mapped to the hardware qubits $Q_3$ and $Q_4$. 
        Because these qubits are adjacent in the topology, the gate can now be executed natively.
    \end{example}

    While straightforward to implement, this local routing strategy often incurs significant gate overhead~\cite{zulehnerEfficientMethodologyMapping2019}. 
    Unfortunately, finding a mapping that globally minimizes the total number of additional SWAP gates is known to be NP-hard~\cite{siraichi_qubit_2019, botea_complexity_2021}. 
    To mitigate this, a variety of heuristic search methods have been developed~\cite{zulehnerEfficientMethodologyMapping2019, sabre, zou_lightsabre_2024, cowtanQubitRoutingProblem, tang2024alpharouterquantumcircuitrouting}. 
    The present work focuses on the A*-based mapping algorithm introduced by Zulehner et al.~\cite{zulehnerEfficientMethodologyMapping2019}. 
    The following paragraphs outline the algorithm’s details. 

   In this approach, the circuit is typically partitioned into layers of disjoint two-qubit gates. 
   The A* algorithm is then used to find a sequence of SWAP gates that satisfies the connectivity constraints for the current layer while considering the impact on subsequent gates.

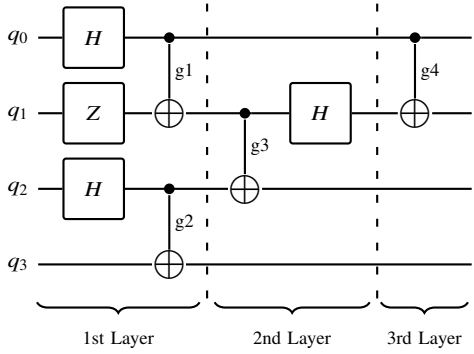
\begin{figure}[h!]
        \centering
        \input{figures/layers}
        \caption{A quantum circuit divided into three layers.}
        \label{fig:layers}
    \end{figure}

    \begin{example}
        \autoref{fig:layers} illustrates a quantum circuit partitioned into three layers based on gate dependency constraints. 
        The first layer comprises two-qubit gates g1 and g2, which can be executed concurrently because they act on disjoint sets of qubits. 
        Because g3 requires the outputs of both g1 and g2, it is assigned to the second layer. 
        Similarly, g4 is placed in the third layer because it depends on g3. Because for fidelity-unaware routing single qubit gates are irrelevant, they can be safely ignored when layering. 
    \end{example}

    As the number of qubits and consequently the concurrent two-qubit gates increases, finding an optimal SWAP sequence within the exponentially large search space becomes increasingly intractable. 
    To maintain feasible runtimes, the mapping algorithm may partition a single layer into multiple sub-layers.

    \begin{example}
    The \enquote{individual gate} strategy decomposes a layer into its two-qubit gates. 
    Consequently, each resulting layer contains only a single gate. 
    For instance, the first \enquote{individual gate} layer in \autoref{fig:layers} is either $\{(q_0,q_1)\}$ or $\{(q_2, q_3)\}$
    \end{example}

    Each node s in the A* search space represents a specific program-to-hardware mapping $\pi_s$:
    \begin{equation}
        \pi_{s}: {q_0, q_1, \ldots, q_{n-1}} \rightarrow {Q_0, Q_1, \ldots, Q_{m-1}}
    \end{equation}
    where $n$ and $m$ denote the number of program and hardware qubits, respectively. We distinguish two types of special nodes:
    \begin{itemize}
        \item \textbf{Root Node}: Represents the initial mapping at the start of the current layer's optimization. It serves as the origin for the search.
        \item \textbf{Goal Node}: A node is defined as a goal node if its associated mapping $\pi_{s}$ satisfies the connectivity constraints for all gates in the current layer.
    \end{itemize}
    The search algorithm terminates upon the discovery of a goal node, yielding a mapping that allows the native execution of all two-qubit operations in the layer.

    The total cost $f(s)$ is determined by the standard A* evaluation function:
    \begin{equation} 
    f(s) = g(s) + h_i(s) 
    \end{equation} 
    Here, $g(s)$ represents the path cost from the root to the current node $s$, defined as the number of SWAP gates applied to reach the current mapping $\pi_s$: 
    \begin{equation} 
    g(s) = \text{dist}(\text{root}, s) 
    \end{equation} 
    The heuristic function $h_{i}(s)$ estimates the remaining cost to satisfy the connectivity requirements of the current layer $L_{i}$. 
    It is calculated as the sum of the distances between the hardware qubits assigned to each program qubit pair in the layer: 
    \begin{equation}
    h_i(s) = \sum_{(q_a, q_b) \in L_i} \left( D(\pi_s(q_a), \pi_s(q_b)) - 1 \right) 
    \end{equation} 
    where the distance function $D(Q_j,Q_k)$ denotes the shortest path length between hardware qubits $Q_j$ and $Q_k$ in the coupling graph $G$.

    Starting from the root, the algorithm iteratively expands the node with the lowest total cost $f(s)$ until a goal node is reached. To expand a node $s$, the algorithm identifies all two-qubit gates within the current layer $L_i$ and their mapped hardware qubits. For each such gate, the algorithm generates child nodes by applying all possible SWAP operations that involve at least one of these hardware qubits. Consequently, each child corresponds to a unique mapping $\pi_{s}$ resulting from a single SWAP. These new states are then inserted into a minimum priority queue — the frontier — for fast retrieval. 

    \begin{figure}[h!]
        \centering
        \input{figures/expansion}
        \caption{The considered SWAP gates on the architecture shown in~\autoref{fig:novera} for a layer consisting solely of the gate $(q_0, q_1) \mapsto (Q_6, Q_1)$.}
        \label{fig:expansion}
    \end{figure}
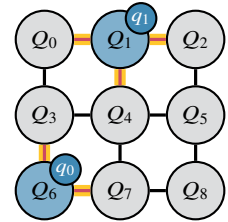

\begin{figure}[h!]
        \centering
        \input{figures/astar}
        \caption{The first two expansion steps of the A* search algorithm for a layer consisting solely of the gate $(q_0, q_1) \mapsto (Q_6, Q_1)$.}\label{fig:astar}
    \end{figure}
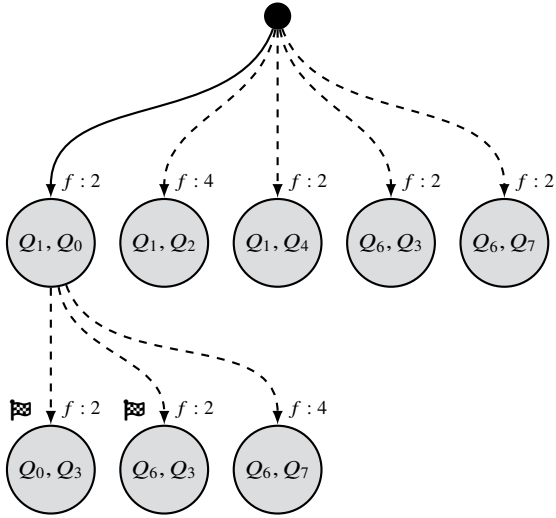

    \begin{example}
        \autoref{fig:astar} illustrates the first two expansion steps of the A* search for a layer  consisting of a single gate: $(q_0, q_1) \mapsto (Q_6, Q_1)$. 
        In the initial expansion, the algorithm generates successor nodes and evaluates the cost function $f(s)$ for the root node for all feasible SWAP operation. 
        \autoref{fig:expansion} illustrates the process of finding feasible SWAP operations for the gate $(q_0, q_1) \mapsto (Q_6, Q_1)$. 
        The algorithm considers all SWAP operations incident to either $Q_1$ or $Q_6$.
        For hardware qubit $Q_1$, the available operations are $(Q_0,Q_1)$, $(Q_1,Q_4)$, and $(Q_1,Q_2)$,. 
        Similarly, for $Q_6$, the feasible SWAP gates are $(Q_6,Q_7)$ and $(Q_6,Q_3)$.
        After the initial expansion, the algorithm selects the node with the lowest total cost for the next expansion. 
        If multiple successors have identical costs, we choose one at random. 
        In the second expansion step, the search proceeds by evaluating a new set of potential SWAPs based on the updated program-to-hardware mapping, and continues this iterative process until a goal state is visited.
        In this example, we generate two potential goal nodes in the second expansion step: $(Q_0,Q_3)$ and $(Q_6, Q_3)$.
        Consequently, the algorithm terminates, if it were to visit one of these nodes.
    \end{example}

    Because the SWAP sequence and the resulting layout thereof directly influence subsequent layers, the heuristic may also consider future layers. The \emph{lookahead-aware} heuristic cost function is defined as:
    \begin{equation}
      \hat{h_{i}}(s) = \sum_{j\in{}W(i,\,N_{L})} \lambda^{j - i} \cdot h_{j}(s),
    \end{equation}
    where $W(j, N_{L}) = (k)_{j \leq k < j + N_{L}}$ and the weight $\lambda \in (0, 1)$ decays exponentially to decrease the contribution of the lookahead layers.

    \begin{example}
        The initial \enquote{front} of the circuit shown in~\autoref{fig:layers} is the layer $\{(q_0, q_1), (q_2, q_3)\}$.
        The goal of the A* search is to find a SWAP sequence such that both of these gates are executable on the target architecture.
        Given the window size $N_L = 3$ and the weight $\lambda = 0.5$, the lookahead layers for this front are $\{(q_1, q_2)\}$ and $\{(q_0, q_1)\}$ with decay factors $0.5$ and $0.25$, respectively. 
        The two-qubit gates inside the lookahead layers contribute to the heuristic cost function $\hat{h_{0}}(v)$ but don't have to be executable, yet.
    
    \end{example}

    \section{Motivation}\label{sec:motivation}

    While MLIR has been leveraged for various tasks in the quantum compilation pipeline, a significant research gap remains in its use for complex, non-local transformations.

    Existing literature features several specialized dialects: the \emph{Quantum} dialect focuses on lowering MLIR to low-level IRs~\cite{mccaskey_mlir_2021}, while \emph{QSSA} introduces static analysis passes to enforce the no-cloning theorem~\cite{peduri_qssa_2022}. 
    Optimization efforts have largely remained localized: For instance, \emph{QIRO} utilizes data flow analysis for local circuit improvements~\cite{ittah_qiro_2022}, and Xanadu’s \emph{Catalyst} framework targets just-in-time (JIT) compilation~\cite{ittah_catalyst_2024}. 
    Furthermore, IBM’s \emph{qe-compiler} employs MLIR to generate executables for quantum control systems~\cite{healy_design_2024}, and the \emph{ASDF} compiler provides a foundation for basis-oriented quantum languages~\cite{adams2025asdfcompilerqwertybasisoriented}. 
    While QLLVM~\cite{zhu2026qllvmscalablequantumclassicalcocompilation} fundamentally relies on MLIR for its compilation flow, it implements the SABRE algorithm at the lower LLVM abstraction level, bypassing the structural advantages of MLIR.

    Collectively, these projects prioritize local optimization routines and translation between abstraction layers. 
    However, a robust, end-to-end quantum compiler requires efficient handling of global optimizations, such as quantum circuit mapping. 
    To date, a purely MLIR-native quantum compilation stack has not been presented in the literature. 
    As a fundamental building block of superconducting qubit compilation, quantum circuit mapping is an ideal candidate for demonstrating that MLIR is well-suited to addressing complex, end-to-end challenges.

    \section{Proposed Solution}\label{sec:proposed-solution}

    While the literature is rich with quantum circuit mapping algorithms and their respective implementations, MLIR-native approaches remain largely unexplored.
    This section details the modular components of the proposed mapping pass for superconducting architectures.
    Each component demonstrates how the strengths of MLIR can be leveraged to build reusable and maintainable solutions without compromising on performance.

    Although these refinements are primarily architectural rather than algorithmic, as we will demonstrate in the evaluation, these improvements — facilitated by the MLIR ecosystem — exert a significant impact on the solution's overall scalability.

    The source code is publicly available at~\href{https://github.com/munich-quantum-toolkit/core}{github.com/munich-quantum-toolkit/core}.

    \subsection{Program Traversal}

    Efficient program traversal is essential for optimization, transpilation, and general compilation routines.
    We combine the linear qubit type of the QCO dialect with MLIR's efficient def-use chain representation to streamline the traversal of quantum programs.
    Specifically, we implement a \emph{bidirectional (qubit) wire iterator} (referred to as \enquote{wire iterator} in the following) that traverses all operations associated with a single qubit while preserving the underlying data-flow dependencies.
    By composing multiple such iterators for the individual qubits in a quantum program, we enable the implementation of complex, multi-qubit traversal patterns and global program analysis.

     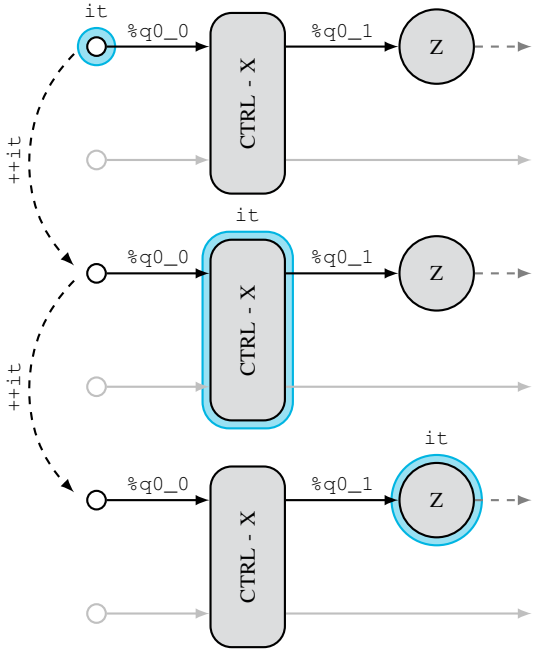
\begin{figure}[h!]
        \input{figures/wire-iterator}
        \caption{Iteration of a wire iterator. The blue highlight indicates the current position of the iterator.}\label{fig:wire-iterator}
    \end{figure}

    \begin{example}
        \autoref{fig:wire-iterator} shows the forward iteration process of the proposed wire iterator.
        The iterator \texttt{it} first points to the allocation operation of the first qubit (indicated by a blue highlight around the white circle).
        After one increment (\texttt{++it}), it points to the controlled operation.
        Incrementing again, it follows the output qubit SSA value of the controlled operation and, afterwards, points to the Z operation.
    \end{example}

    We utilize the wire iterator to divide the program into layers of disjoint two-qubit operations.
    The traversal is initialized with each iterator positioned at the qubit allocation operations.
    During each iteration, the iterators advance along their respective wires until a two-qubit operation is encountered.
    A two-qubit operation is marked as \enquote{ready} and inserted into the current layer only when it has been reached twice---once by each of the two involved qubits' iterators.
    This ensures that all dependencies for the two-qubit operation have been satisfied, and it can be executed in parallel with other operations in the same layer.

\begin{figure}[h!]
        \centering
        \input{figures/layering}
        \caption{Finding the two-qubit gates (highlighted in green) of a layer.}\label{fig:layering}
    \end{figure}
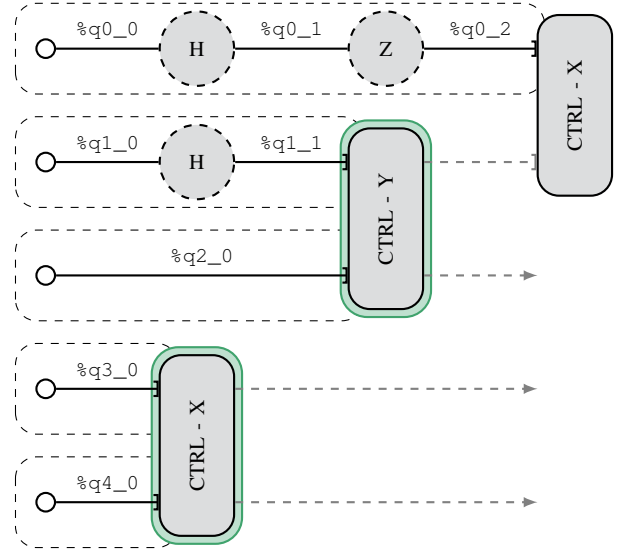

    \begin{example}
        \autoref{fig:layering} illustrates the first iteration of the layer-finding algorithm for five qubits. 
        In this example, the layer consists of two controlled operations, which we highlight in green. 
        The algorithm advances the wire iterator of each qubit until a two-qubit operation is encountered. 
        If two wire iterators reach the same two-qubit operation in the same iteration, the two-qubit operation is considered \enquote{ready} and inserted into the current layer. 
        \enquote{Unready} operations, such as the controlled-X acting on the upper two qubits, belong to one of the subsequent layers.
        Lastly, the wire iterators of the \enquote{ready} two-qubit operations are incremented and the next iteration starts.
    \end{example}
    
    By initializing the wire iterators at the qubit deallocation operations and decrementing them, the same logic can be applied to traverse the circuit backwards.
    This symmetry allows for an analogous backward traversal to compute reverse-order layers, which can be beneficial for certain optimization strategies that consider both forward and backward passes through the program, such as the SABRE algorithm~\cite{sabre}.

    To further minimize the number of layers, the algorithm aggregates operations into two-qubit blocks.
    We define such a block as the sequence of all single- and two-qubit operations acting on the same pair of qubits as a \enquote{defining} two-qubit operation.
    If this defining two-qubit operation satisfies the target architecture's connectivity constraints, all operations within the block are guaranteed to be executable as well.
    Leveraging MLIR's def-use chain, this strategy is efficiently implemented by advancing a pair of wire iterators in tandem until they encounter an operation involving a third qubit.

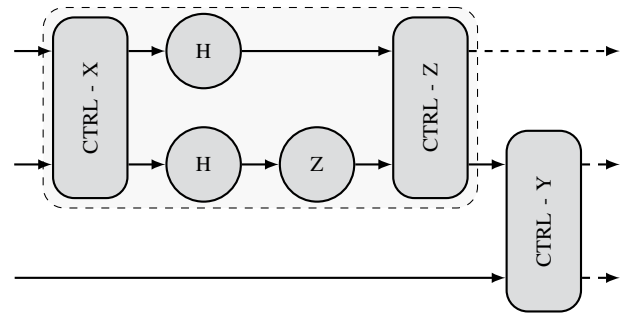
\begin{figure}[h!]
        \centering
        \input{figures/two-qubit-block}
        \caption{A two-qubit block aggregates all one and two-qubit operations acting on the same wires.}\label{fig:two-qubit-block}
    \end{figure}

    \begin{example}
        \autoref{fig:two-qubit-block} shows the composition of a two-qubit block.
        The two-qubit block starts with the controlled-X operation on the upper two qubits and contains all single- and two-qubit operations acting on the same qubits.
        In this example, these operations are the two Hadamards, the Z operation, and finally, the controlled-Z operation. The controlled-Y operation does not belong to the two-qubit block because it acts on a third qubit, namely the lowest one.
    \end{example}
    
    Finally, we enable program traversal across arbitrary gate sets by leveraging the \texttt{UnitaryOpInterface} within the QCO dialect. 
    While existing approaches~\cite{wille2023qmap} are often constrained by hard-coded gate sets, the proposed solution remains entirely gate-agnostic. 
    This design future-proofs the mapping pass against evolving architectures and varying native gate-sets.

    \subsection{Arena-Based A* Search}

    As the most computationally intensive component of the mapping pass, the performance of the A* search is integral to the overall performance.
    By utilizing MLIR’s built-in data structures and memory management primitives, we introduce several architectural refinements over prior implementations.

    The implementation of the A* mapping algorithm in~\cite{wille2023qmap} stores the complete sequence of SWAP operations from the root to the current node for each node of the A* search graph. This design imposes a significant memory overhead, as each child node redundantly replicates its parent's path history. To mitigate this, we propose an arena-based A* search. Instead of storing the full path, each node maintains only a pointer to its parent, allowing the optimal SWAP sequence to be reconstructed once a goal node is found. To manage these nodes, we leverage MLIR’s \texttt{SpecificBumpPtrAllocator}, which allocates nodes in contiguous memory blocks by simply advancing a pointer within a pre-allocated (possibly growing) memory region -- the \enquote{arena}. Usually, the frontier manages and stores the nodes of the search graph. Because in the proposed solution the arena contains the nodes, the frontier consists of node pointers instead. This approach avoids memory fragmentation with near-zero allocation overhead and allows the entire search graph to be deallocated in a single operation upon completion.

     \begin{figure}[h!]
        \centering
        \input{figures/reconstruction}
        \caption{Reconstructing the SWAP sequence in the search-graph via parent pointers.}
        \label{fig:reconstruction}
    \end{figure}
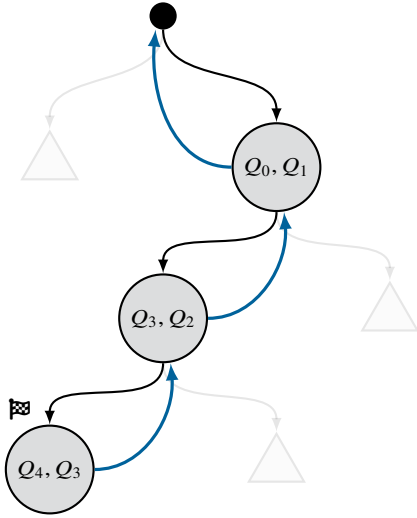
    
    \begin{example}
        \autoref{fig:reconstruction} illustrates the SWAP sequence reconstruction mechanism. The reconstruction originates at the goal node (indicated by the flag symbol) and traverses the solution path via parent pointers until the root node is reached. In the example shown, the sequence $(Q_4,Q_3)$, $(Q_3,Q_2)$, and finally $(Q_0, Q_1)$ is extracted. This sequence is subsequently reversed to yield the final path from the root to the goal node.
    \end{example}

    We minimize reliance on the C++ Standard Library (\texttt{std::}) in favor of specialized data structures from the MLIR ecosystem. For instance, we utilize \texttt{SmallVector} to reduce heap allocations by leveraging its inline storage for small-to-moderate element counts. Similarly, replacing standard maps with LLVM-native associative containers, such as \texttt{DenseMap}, further optimizes cache locality. Moreover, \texttt{llvm::PriorityQueue} serves as a drop-in replacement for \texttt{std::priority\_queue} while providing convenience functions to re-order elements efficiently.

    Because distinct SWAP sequences can yield the same program-to-hardware qubit mapping, expanding nodes with identical states introduces significant search redundancy.
    To prune suboptimal paths, we implement a closed-set mechanism by storing these mappings in an MLIR \texttt{DenseMap}, which maintains the minimum SWAP sequence length for each mapping.
    A node is only expanded if its number of SWAPs is strictly less than the previously seen one.
    
\begin{figure}[h!]
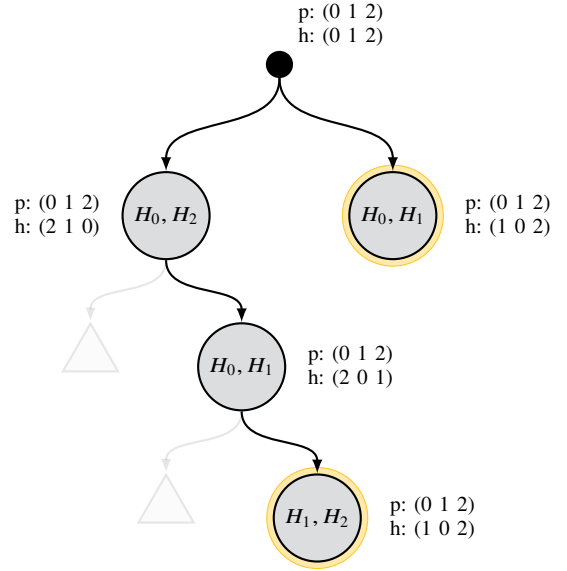

        \centering
        \include{figures/same-layout-tree.tex}
        \caption{Two paths in the search graph lead to the same program-to-hardware qubit mapping.}\label{fig:same-layouts-tree}
    \end{figure}

    \begin{example}
        \autoref{fig:same-layouts-tree} illustrates a scenario where distinct paths within the search space yield the same program-to-hardware qubit mapping. The search starts at the root node with an identity mapping, where each program qubit $q_{i}$ is assigned to its corresponding hardware qubit $Q_{i}$. As shown, the SWAP sequences $((0,1))$ and $((0,2),(0,1),(1,2))$ result in the same qubit permutation: $(0,1,2) \mapsto (1,0,2)$. Expanding these equivalent nodes (highlighted in blue) yields identical sub-trees, as the expanding strategy depends solely on the current mapping rather than the path taken to reach it.
    \end{example}

    \subsection{Parallel Initial Layout Refinement}

    The initial program-to-hardware qubit mapping substantially influences the total SWAP overhead of the final circuit. To address this, prior work~\cite{sabre} introduced an iterative bidirectional strategy that refines the initial mapping through successive forward and backward routing passes. The authors in~\cite{zou_lightsabre_2024} augment this idea by executing multiple independent "trials" in parallel to explore a broader range of initial configurations. Finally, the best candidate is chosen based on the number of inserted SWAPs. We adapt these strategies for the proposed A*-based mapping algorithm to explore different solution paths in the search-graph.
    
    Unlike original SABRE-based implementations~\cite{sabre, zou_lightsabre_2024}, which generate a reconstruction plan during the final pass to rebuild the program, the proposed approach inserts SWAP operations directly into the IR. As a consequence, counting the number of inserted SWAPs during the final forward pass requires deep copying the IR for each thread.
    To maintain memory efficiency during parallel execution, we instead collect the number of inserted SWAPs during the final backward pass, therefore choosing the best initial layout for the final forward pass heuristically.

    Whereas high-performance parallelism often requires third-party dependencies such as OpenMP, MLIR mitigates this overhead by providing native multithreading support and exposing primitives for parallel computation. We leverage these utilities to implement the parallel trial strategy. The \texttt{parallelForEach} function implements task-based parallelism by executing a  callback function on a collection of initial configurations. The following snippet shows the implementation of this concurrent execution logic within the mapping pass, where \texttt{ntrials} is the number of random trials and \texttt{nqubits} is the number of qubits the architecture supports.

        \begin{Verbatim}[fontsize=\footnotesize, frame=lines, framesep=2mm, baselinestretch=1.2]
// Generate `ntrials` many random layouts:
SmallVector<Trial> trials;
trials.reserve(ntrials);
for (std::size_t i = 0; i < trials.size(); ++i) {
    trials.emplace_back(
        Layout::random(nqubits, seed);
    );
}

// Run the trials in parallel:
mlir::parallelForEach(&getContext(), trials,
    [&, this](Trial& trial) {
        refineLayout(func, arch, trial);
    });
    \end{Verbatim}

    \subsection{SWAP Insertion}

    After selecting the most promising refined initial layout, the mapping pass performs a final \enquote{hot} routing phase by directly inserting SWAP operations in the IR.
    This strategy avoids the computational overhead of reconstructing the IR from scratch and the maintenance of auxiliary data structures typically required for a full rebuild.

    To insert a SWAP operation into the IR, we utilize the MLIR \texttt{IRRewriter}. 
    This process involves setting the insertion point through the rewriter's API and instantiating the operation via \texttt{SWAPOp::create}. 
    Because a SWAP operation exchanges the states of its two input qubits, the pass must rewire the output SSA values to maintain the intended program semantics.

    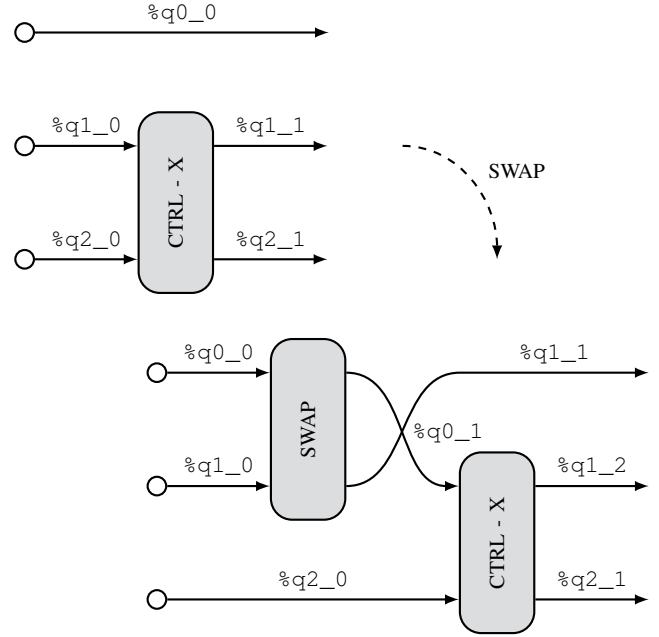
\begin{figure}[h!]
        \centering
        \input{figures/rewire}
        \caption{The rewiring process after SWAP insertion maintains the intended program semantics.}
        \label{fig:rewire}
    \end{figure}
    \begin{example}
       \autoref{fig:rewire} illustrates the rewiring process. 
       The inserted SWAP operation exchanges the states of the input qubits \texttt{\%q0\_0} and \texttt{\%q1\_0}. 
       Consequently, to maintain semantic equivalence, the operands of the controlled-X operation are updated such that it now consumes \texttt{\%q0\_1} as its input.
    \end{example}

    However, inserting SWAP operations solely based on data-flow dependencies may result in a structurally invalid program, as the operations that produce the input SSA values are not guaranteed to precede the SWAP in the IR’s linear order. 
    If such an SSA dominance violation occurs, the pass fails, and subsequent passes in the pass pipeline are aborted. 
    To resolve this, we perform a final topological sort of the program using the \texttt{sortTopologically} utility provided by MLIR.

    \begin{example}
       The following snippet illustrates an invalid IR (omitting allocations, sinks, and types for clarity). 
       The highlighted SWAP operation consumes SSA values as operands that are defined later in the instruction sequence. 
       This out-of-order definition violates SSA dominance, which requires that a value be defined before it is used.
    \end{example}

    \begin{Verbatim}[fontsize=\footnotesize, frame=lines, framesep=2mm, baselinestretch=1.2, highlightlines={3}, highlightcolor=IEEECyan!10]
%q2_1, %q1_1 = qco.ctrl(%q2_0) 
                   targets (%arg0 = %q1_0) { ... }
%q5_2, %q3_1 = qco.swap %q5_1, %q3_0
%q1_2, %q0_1 = qco.ctrl(%q1_1)
                   targets (%arg0 = %q0_0) { ... }
%q3_2, %q2_2 = qco.ctrl(%q3_1) 
                   targets (%arg0 = %q2_1) { ... }
%q5_1, %q4_1 = qco.ctrl(%q5_0) 
                   targets (%arg0 = %q4_0) { ... }
    \end{Verbatim}

     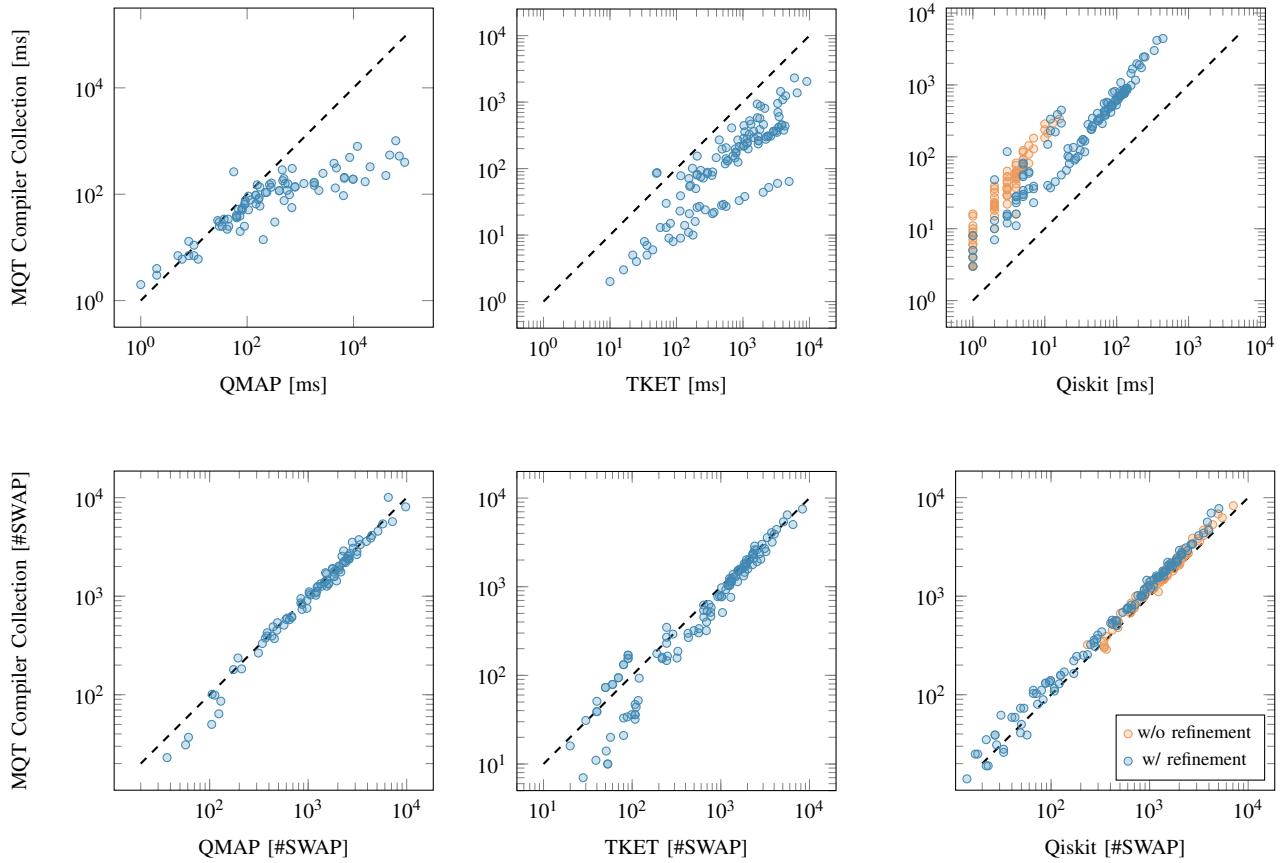
\begin{figure*}[t]
        \centering
        \begin{subfigure}[t]{0.32\textwidth}
            \centering
            \input{plots/qmap-comparison-runtime}
        \end{subfigure}%
        \begin{subfigure}[t]{0.32\textwidth}
            \centering
            \input{plots/tket-comparison-runtime}
        \end{subfigure}%
        \begin{subfigure}[t]{0.32\textwidth}
            \centering
            \input{plots/qiskit-comparison-runtime}
        \end{subfigure}

        \vspace{1em}
        
        \begin{subfigure}[t]{0.32\textwidth}
            \centering
            \input{plots/qmap-comparison-nswaps}
        \end{subfigure}%
        \begin{subfigure}[t]{0.32\textwidth}
            \centering
            \input{plots/tket-comparison-nswaps}
        \end{subfigure}%
        \begin{subfigure}[t]{0.32\textwidth}
            \centering
            \input{plots/qiskit-comparison-nswaps}
        \end{subfigure}
        
        \caption{A comparison between QMAP, TKET, Qiskit, and the proposed solution.}
        \label{fig:comparison}
    \end{figure*}

    \newpage

    \section{Evaluation}\label{sec:evaluation}
    
    As one of the most computationally complex tasks in an end-to-end quantum compilation pipeline, the runtime and solution quality plays an integral part of the overall performance of the compiler. 
    Thus, this section evaluates the proposed solution and compares it to state-of-the-art implementations. 

    \subsection{Experimental Setup}

   To ensure the reproducibility of our results, this section details the environment used for all benchmarks. 
   We conducted our experiments on an Apple M5 Pro (18-core CPU) equipped with 24 GB of RAM running macOS Tahoe 26.3.

    By implemented the benchmarks as a standalone C++20 executable, we eliminate the overhead of Python bindings and ensure peak performance. 
    The project was compiled using Clang 22.1.3. 
    Detailed version information for all external libraries can be found in the table below. 
    To further optimize the MLIR framework's performance, we performed a local build with assertions disabled.

    \begin{table}[h!]
    \centering
        \begin{tabular}{l c}
            \hline
            \textbf{Library} & \textbf{Version} \\ 
            \hline
            QMAP & 2.6.0 \\
            TKET & 2.1.77 \\
            Qiskit & 2.4.0 \\
            MLIR (LLVM) & 22.1.3 \\
            \hline
        \end{tabular}
    \end{table}

    We target a 10$\times$12 square lattice architecture for the mapping pass across all benchmarks, modeling IBM’s state-of-the-art 120-qubit Nighthawk processor~\cite{IBMNighthawk2026}. 
    To evaluate the proposed solution under production-grade conditions, we benchmark almost hundred quantum programs with sizes ranging from 30 to 120 qubits. 
    We specifically focus on the upper end of this range—the \enquote{utility-scale} spectrum to demonstrate scalability.
    These benchmarks are generated using MQT Bench~\cite{quetschlichMQTBenchBenchmarking2023} and the results are shown in~\autoref{fig:comparison}.

    \subsection{MQT Compiler Collection vs. QMAP}

    Both the QMAP library and the proposed solution implement the same A* mapping algorithm. 
    However, while QMAP utilizes bespoke data structures, our solution is built upon the MLIR framework. 
    The following results evaluate how leveraging MLIR as a compilation backbone impacts both runtime performance and solution quality.

    For benchmarking, both implementations use an individual-gate strategy with 15 lookahead layers, and the proposed solution also incorporates two-qubit block aggregation. 
    Regarding initial placement, QMAP employs a single bidirectional iteration (a forward and backward pass) on its dynamic layout, whereas our approach evaluates 18 random initial layouts to more broadly explore the search space. Both algorithms apply a decay factor $\lambda$ of 0.5.

    On average, the proposed solution reduces SWAP insertions by 5\,\% and runtime by 46\,\%. The SWAP reduction is primarily due to the parallel random layout trials, which offer a more thorough search than QMAP’s dynamic strategy. Notably, the 46\,\%. runtime improvement underscores the efficiency of the MLIR ecosystem, demonstrating that the A* algorithm can be significantly accelerated by leveraging MLIR primitives and optimized data structures.

    \subsection{MQT Compiler Collection vs. TKET}

    The TKET library offers state-of-the-art quantum compilation primitives including a specialized quantum circuit mapping algorithm~\cite{Sivarajah_TKET_A_Retargetable_2020}. 
    This section demonstrates that by leveraging the MLIR framework, the proposed solution outperforms TKET's \texttt{LexiRoute}~\cite{cowtanQubitRoutingProblem} method in terms of runtime and SWAP overhead.

    In our experimental setup, TKET was configured to utilize its default placement strategy alongside the \texttt{LexiRouteRoutingMethod} for circuit routing.
    Driven by the significant runtime improvements observed in the previous benchmarks, we increased the lookahead depth of the proposed solution to 32 to further optimize the mapping quality.
    To prevent the decay factor from vanishing over this extended horizon, the decay parameter $\lambda$ was adjusted to 0.7.

    The proposed solution achieves a 13\,\% reduction in the number of inserted SWAP gates. 
    While the search algorithm is a primary driver of this improvement, the dynamic initial layout strategy is similar to TKET's default, and therefore, this reduction may be partially attributed to the random trial mechanism. 
    Furthermore, the implementation requires, on average, 74\,\% less runtime to compute and insert these SWAPs, surpassing the runtime improvements observed in the previous QMAP benchmark. 
    These results highlight the efficiency of the A* search mapping algorithm and underscore the performance advantages of its MLIR-native implementation.

    \subsection{MQT Compiler Collection vs. Qiskit}

    The final evaluation compares the proposed solution to Qiskit’s SABRE, a leading algorithm widely regarded as a state-of-the-art benchmark for circuit mapping.

    The configuration of our solution remains identical to that in the previous section, except for the iteration count, which was increased to 4 to match SABRE’s default setting. 
    Both implementations were evaluated using 18 parallel trials.

    To isolate the impact of layout refinement strategies, we conducted an additional benchmark comparing runtime and solution quality across a single forward pass using an identical, randomly selected initial layout for both tools.

    The results indicate that SABRE outperforms the proposed solution in both metrics. 
    Specifically, SABRE inserts 8\,\% fewer SWAPs and requires 28\,\% less runtime in the bidirectional benchmark. 
    While our solution identifies better mappings (fewer SWAPs) in over one-third of the test cases, the results in the non-refined benchmark ultimately confirm SABRE's algorithmic advantage.

    In summary, these results suggest that while the MLIR framework can significantly enhance the performance of existing mapping implementations, it is not a silver bullet. 
    It cannot magically bridge the gap when compared to fundamentally more efficient algorithmic designs.

    \section{Conclusions}\label{sec:conclusion}

    This paper presents a significant step toward an MLIR-native, end-to-end quantum compilation pipeline by evaluating the framework's capacity for global optimization. 
    We have demonstrated that by leveraging the MLIR ecosystem and its built-in primitives, it is possible to achieve substantial improvements in both runtime performance and solution quality compared to tailored implementations such as QMAP and TKET. 
    These results underscore the performance advantages of integrating quantum primitives within a robust, general-purpose compiler framework.

    However, as our comparison with Qiskit's SABRE illustrates, a high-performance foundation like MLIR complements rather than substitutes for sophisticated algorithmic design and theoretical complexity analysis: The underlying mapping logic remains the ultimate arbiter of success.

    Future research could extend these concepts by incorporating hardware calibration data into the search algorithm to enable noise-aware quantum circuit mapping. 
    Furthermore, to support the growing demand for hybrid quantum-classical computing, subsequent iterations should incorporate classical control flow. 
    Finally, leveraging MLIR’s graph regions to model these hybrid programs offers a promising avenue to eliminate the computational overhead of topological sorting after SWAP insertion. 

    The MQT Compiler Collection is an open-source project and publicly available at~\href{https://github.com/munich-quantum-toolkit/core}{github.com/munich-quantum-toolkit/core}.

    \ifshowauthors

    \subsection*{Acknowledgments}\label{sec:ack}

    This work received funding from the European Research Council (ERC) under the European Union’s Horizon 2020 research and innovation program (grant agreement No. 101001318), was part of the Munich Quantum Valley, which the Bavarian state government supports with funds from the Hightech Agenda Bayern Plus, and has been supported by the BMK, BMDW, and the State of Upper Austria in the frame of the COMET program (managed by the FFG).

    While preparing the manuscript, Anthropic's Claude, OpenAI's GPT, and Google's Gemini were used to improve readability, spelling, grammar, and clarity throughout the manuscript. Each LLM output was reviewed by the authors and edited manually as needed. The authors take full responsibility for the final content.
    
    \else
    \fi

    \clearpage
    \printbibliography

\end{document}

%% file: preamble.tex
\usepackage[T1]{fontenc} 
\usepackage[utf8]{inputenc} 
\usepackage[english]{babel} 
\usepackage{amsmath,amsthm,amsfonts,amssymb} 
\usepackage{colonequals} 
\usepackage[cmintegrals]{newtxmath} 
\usepackage{mathtools} 
\usepackage{thmtools} 
\usepackage{stmaryrd} 
\usepackage{bm} 
\usepackage{siunitx} 
\usepackage{comment} 
\usepackage{microtype} 
\usepackage{fnpct} 
\usepackage{fontawesome5} 
\usepackage{csquotes} 
\usepackage[inline]{enumitem} 
\usepackage{tabularx} 
\usepackage{booktabs} 
\usepackage[hidelinks]{hyperref} 
\usepackage[capitalize,nameinlink]{cleveref} 
\usepackage{graphicx} 
\usepackage[inkscapelatex=false]{svg} 
\usepackage{wrapfig} 
\usepackage{float}  
\usepackage{placeins} 
\usepackage{xspace} 
\usepackage[noEnd]{algpseudocodex} 
\usepackage[
    backend=biber,
    style=ieee,
    sortlocale={en_US},
    sortcites=true,
    url=true,
    eprint=true,
    giveninits=false,
    citestyle=numeric-comp,
    minnames=1,
    maxnames=5
]{biblatex}

\usepackage{tikz}
\usetikzlibrary{arrows,decorations.pathreplacing,fit, fadings, backgrounds, shapes.geometric, shadows, calc, topaths, tikzmark, arrows.meta}
\usepackage{fancyvrb}
\makeatletter
\let\MYcaption\@makecaption
\makeatother
\usepackage{subcaption} 
\makeatletter
\let\@makecaption\MYcaption
\makeatother
\AtEveryBibitem{
    \clearname{editor}%
    \clearfield{series}%
    \clearfield{isbn}%
    \clearfield{issn}%
    \clearfield{day}%
    \clearfield{month}%
    \clearfield{place}%
    \clearlist{location}%
    \clearfield{pages}%
    \clearfield{volume}%
    \clearfield{number}%
    \clearfield{eprintclass}%
    \ifentrytype{software}{%
    }{%
        \clearfield{url}%
        \clearfield{urlyear}%
        \clearfield{urldate}%
    }%
}
\DeclareSourcemap{
    \maps[datatype=bibtex]{
        \map{
            \step[fieldsource=doi, match=\regexp{.+/arXiv\..+}, final]
            \step[fieldsource=eprint, match=\regexp{.+}, final]
            \step[fieldset=doi, null]
        }
        \map{
            \step[fieldsource=doi, match=\regexp{.+}, final]
            \step[fieldset=eprint, null]
            \step[fieldset=archiveprefix, null]
            \step[fieldset=eprinttype, null]
        }
    }
}
\graphicspath{ {figures/} }
\setlist[enumerate]{label=(\arabic*)} 
\faStyle{regular}
\declaretheoremstyle[
    bodyfont=\itshape,%
]{example-style}
\declaretheorem[
    name=Example,%
    style=example-style,%
    numbered=unless unique,%
]{example}
\crefname{section}{Sec.}{Sec.}
\Crefname{section}{Sec.}{Sec.}
\crefname{example}{Ex.}{Ex.}
\Crefname{example}{Ex.}{Ex.}

\definecolor{TUM_blue}{RGB}{0,101,189}
\colorlet{TUM_black}{black}
\colorlet{TUM_white}{white}
\definecolor{TUM_darkblue}{RGB}{0,82,147}
\colorlet{TUM_darkblue100}{TUM_darkblue}
\colorlet{TUM_darkblue80}{TUM_darkblue100!80}
\colorlet{TUM_darkblue50}{TUM_darkblue100!50}
\colorlet{TUM_darkblue20}{TUM_darkblue100!20}
\definecolor{TUM_verydarkblue}{RGB}{0,51,89}
\colorlet{TUM_verydarkblue100}{TUM_verydarkblue}
\colorlet{TUM_verydarkblue80}{TUM_verydarkblue100!80}
\colorlet{TUM_verydarkblue50}{TUM_verydarkblue100!50}
\colorlet{TUM_verydarkblue20}{TUM_verydarkblue100!20}
\colorlet{TUM_darkgrey}{TUM_black!80}
\colorlet{TUM_grey}{TUM_black!50}
\colorlet{TUM_lightgrey}{TUM_black!20}
\definecolor{TUM_beige}{RGB}{218,215,203}
\definecolor{TUM_orange}{RGB}{227,114,34}
\definecolor{TUM_green}{RGB}{162,173,0}
\definecolor{TUM_verylightblue}{RGB}{152,198,234}
\definecolor{TUM_lightblue}{RGB}{100,160,200}
\newcommand{\emptylines}[1]{%
    \newcount\foo%
    \foo=0%
    \loop%
    \phantom{}\\%
    \advance\foo 1%
    \ifnum
        \foo<#1%
        \repeat%
}
\ifshowcomments
\usepackage[textwidth=5cm]{todonotes} 
\usepackage[switch]{lineno} 
\linenumbers
\newlength{\widen}
\advance\paperwidth by \dimexpr\widen+\widen\relax
\advance\evensidemargin by \widen
\advance\oddsidemargin by \widen
\marginparwidth=\dimexpr\widen+0mm\relax 
\newcommand{\mytodo}[4]{\bgroup\color{#2!65!black}#3\egroup\todo[size=\scriptsize,color=#2]{#1: #4}}
\else
\newcommand{\mytodo}[4]{#3}
\fi
\definecolor{yannick}{HTML}{e89c1c}
\definecolor{robert}{HTML}{e5dd0c}
\definecolor{lukas}{HTML}{76c11f}
\definecolor{patrick}{HTML}{3fb86e}
\definecolor{matthias}{HTML}{14b292}

\newcommand{\matthias}[2][]{\mytodo{M}{matthias}{#1}{#2}}

\usepackage{hhline} 
\usepackage{colortbl} 
\usepackage[table,xcdraw]{xcolor} 
\definecolor{lightgray}{gray}{0.95} 
\definecolor{myblue}{HTML}{CCDCE9} 

\usepackage{fvextra}
\usepackage{pgfplots}

\definecolor{IEEEBlue}{HTML}{00629B}
\definecolor{IEEECyan}{HTML}{00B5E2}
\definecolor{IEEEOliveGreen}{HTML}{658D1B}
\definecolor{IEEEDarkYellow}{HTML}{FFC72C}
\definecolor{IEEERed}{HTML}{BA0C2F}
\definecolor{IEEEGray}{HTML}{75787B}
\definecolor{IEEEGreen}{HTML}{00843D}
\definecolor{IEEEDarkPurple}{HTML}{772583}
\definecolor{IEEEDarkOrange}{HTML}{E87722}

%% file: figures/dataflow.tex
\tikzset{
    measurement/.pic = {
        \draw[thick] (0.2, 0.25) arc (180:0:0.2);  
        \draw[thick] (0.35, 0.275) -- ++(0.2,0.2); 
    }
}

\begin{tikzpicture}[font=\footnotesize]
    \node[circle, draw=black, thick, minimum size=7pt, inner sep=0pt, fill=white] (q0) at (0, 1.5) {};
    \node[circle, draw=black, thick, minimum size=7pt, inner sep=0pt, fill=white] (q1) at (0, 0) {};

    \node[circle, minimum size=28pt, fill=IEEEGray!25, draw=black, thick] (h) at (1.75, 1.5) {H};

    \node[rectangle, rounded corners=8pt, minimum height=68pt, minimum width=28pt, fill=IEEEGray!25, draw=black, thick] (cx) at (4, 0.75) {\rotatebox{90}{CTRL - X}};

    \node[circle, minimum size=28pt, fill=IEEEGray!25, draw=black, thick] (measq0) at (6.5, 1.5) {};
    \pic at (6.1, 1.15) {measurement};

    \node[circle, minimum size=28pt, fill=IEEEGray!25, draw=black, thick] (measq1) at (6.5, 0) {};
    \pic at (6.1, -0.35) {measurement};

    \node[circle, draw=black, thick, minimum size=7pt, inner sep=0pt, fill=black] (q0end) at (8, 1.5) {};
    \node[circle, draw=black, thick, minimum size=7pt, inner sep=0pt, fill=black] (q1end) at (8, 0) {};

    \draw[-latex, thick] (q0.east) -- (h.west) node[midway, above] {\texttt{\%q0\_0}};
    \draw[-latex, thick] (h.east) -- ([yshift=7.5mm]cx.west) node[midway, above] {\texttt{\%q0\_1}} node[below, midway] {\tiny \texttt{ctrl}};;
    \draw[-latex, thick] ([yshift=7.5mm]cx.east) -- (measq0.west) node[midway, above] {\texttt{\%q0\_2}};
    \draw[-latex, thick] (measq0.east) -- (q0end.west) node[midway, above] {\texttt{\%q0\_3}};
    
    \draw[-latex, thick] (q1.east) -- ([yshift=-7.5mm]cx.west) node[midway, above] {\texttt{\%q1\_0}} node[below, xshift=-6.5mm] {\tiny \texttt{target}};
    \draw[-latex, thick] ([yshift=-7.5mm]cx.east) -- (measq1.west) node[midway, above] {\texttt{\%q1\_1}};
    \draw[-latex, thick] (measq1.east) -- (q1end.west) node[midway, above] {\texttt{\%q1\_2}};

    \path (measq0.north) to[out=90, in=180] coordinate[pos=0.5] (c0mid) ++(1, 0.33);
    \draw[-latex, thick, dashed] (measq0.north) to[out=90, in=180] ++(1, 0.33);
    \node[above=2pt, sloped, inner sep=1pt] at (c0mid) {\texttt{\%c0}};

    \path (measq1.south) to[out=260, in=180] coordinate[pos=0.5] (c1mid) ++(1, -0.33);
    \draw[-latex, thick, dashed] (measq1.south) to[out=260, in=180] ++(1, -0.33);
    \node[below=2pt, sloped, inner sep=1pt] at (c1mid) {\texttt{\%c1}};
\end{tikzpicture}

%% file: figures/modifier.tex
\begin{tikzpicture}[font=\footnotesize]
    \node[rectangle, rounded corners=8pt, minimum height=68pt, minimum width=28pt, fill=IEEEGray!25, draw=black, thick] (cx) at (0, -0.25) {\rotatebox{90}{CTRL - X}};

    \draw[latex-, thick] ([yshift=7.5mm]cx.west) -- ++(-1, 0) node[above, midway] {\texttt{\%q0\_1}} node[below, midway] {\tiny \texttt{ctrl}};;
    \draw[latex-, thick] ([yshift=-7.5mm]cx.west) -- ++(-1, 0) node[above, midway] {\texttt{\%q1\_0}} node[below, midway] {\tiny \texttt{target}};;

    \draw[-latex, thick] ([yshift=7.5mm]cx.east) -- ++(1, 0) node[above, midway] {\texttt{\%q0\_2}};
    \draw[-latex, thick] ([yshift=-7.5mm]cx.east) -- ++(1, 0) node[above, midway] {\texttt{\%q1\_1}};

    \node at (2, -0.25) {=};

    \node[rectangle, rounded corners=10pt, minimum height=80pt, minimum width=80pt, fill=IEEEGray!25, draw=black, thick] (cx-modifier) at (5, -5pt) {};

    \node[rectangle, rounded corners=8pt, minimum height=40pt, minimum width=70pt, fill=IEEEGray!10, draw=black, thick] (cx-modifier-target-area) at (5, -20pt) {};

    \node[fill=IEEEDarkYellow!50, draw=IEEEDarkYellow, text=black, rounded corners=4pt] (arg0) at (cx-modifier-target-area.west) {\scriptsize\rotatebox{90}{\texttt{\%arg0}}};

    \node[circle, minimum size=28pt, fill=IEEEGray!25, draw=black, thick] (cx-modifier-x) at (5, -0.7) {X};

    \draw[thick, dashed] ([yshift=7.5mm]cx-modifier.west) -- ([yshift=7.5mm]cx-modifier.east);

    \draw[latex-, thick] ([yshift=7.5mm]cx-modifier.west) -- ++(-1, 0) node[above, midway] {\texttt{\%q0\_1}} node[below, midway] {\tiny \texttt{ctrl}};
    \draw[latex-, thick] (arg0.west) -- ++(-0.95, 0) node[above, midway] {\texttt{\%q1\_0}} node[below, midway] {\tiny \texttt{target}};;

    \draw[thick, dashed] (arg0.east) -- (cx-modifier-x.west);
    \draw[thick, dashed] (cx-modifier-x.east) -- ([yshift=-5.5mm]cx-modifier.east);
    
    \draw[-latex, thick] ([yshift=7.5mm]cx-modifier.east) -- ++(1, 0) node[above, midway] {\texttt{\%q0\_2}};
    \draw[-latex, thick] ([yshift=-5.5mm]cx-modifier.east) -- ++(1, 0) node[above, midway] {\texttt{\%q1\_1}};
\end{tikzpicture}

%% file: figures/novera.tex
\begin{tikzpicture}[scale=1,font=\footnotesize]
\node[circle, draw=black, fill=IEEEGray!25, text=black, thick] (q0) at (-1, 1) {$Q_{0}$};
\node[circle, draw=black, fill=IEEEGray!25, text=black, thick, minimum size=3mm] (q1) at (0, 1) {$Q_{1}$};
\node[circle, draw=black, fill=IEEEGray!25, text=black, thick, minimum size=3mm] (q2) at (1, 1) {$Q_{2}$};

\node[circle, draw=black, fill=IEEEGray!25, text=black, thick, minimum size=5mm] (q3) at (-1, 0) {$Q_{3}$};
\node[circle, draw=black, fill=IEEEGray!25, text=black, thick, minimum size=5mm] (q4) at (0, 0) {$Q_{4}$};
\node[circle, draw=black, fill=IEEEGray!25, text=black, thick, minimum size=5mm] (q5) at (1, 0) {$Q_{5}$};

\node[circle, draw=black, fill=IEEEGray!25, text=black, thick, minimum size=5mm] (q6) at (-1, -1) {$Q_{6}$};
\node[circle, draw=black, fill=IEEEGray!25, text=black, thick, minimum size=5mm] (q7) at (0, -1) {$Q_{7}$};
\node[circle, draw=black, fill=IEEEGray!25, text=black, thick, minimum size=5mm] (q8) at (1, -1) {$Q_{8}$};

\draw[very thick, color=black] (q0.east) -- (q1.west);
\draw[very thick, color=black] (q1.east) -- (q2.west);

\draw[very thick, color=black] (q3.east) -- (q4.west);
\draw[very thick, color=black] (q4.east) -- (q5.west);

\draw[very thick, color=black] (q6.east) -- (q7.west);
\draw[very thick, color=black] (q7.east) -- (q8.west);

\draw[very thick, color=black] (q0.south) -- (q3.north);
\draw[very thick, color=black] (q3.south) -- (q6.north);

\draw[very thick, color=black] (q1.south) -- (q4.north);
\draw[very thick, color=black] (q4.south) -- (q7.north);

\draw[very thick, color=black] (q2.south) -- (q5.north);
\draw[very thick, color=black] (q5.south) -- (q8.north);
\end{tikzpicture}

%% file: figures/mapping.tex
\begin{tikzpicture}[scale=1.0, font=\footnotesize]
    \node[text=black] (q0) at (0, 3.75) {$Q_{4} \mapsfrom q_{0}$};
    \node[text=black] (q1) at (0, 3) {$Q_{1} \mapsfrom q_{1}$};
    \node[text=black] (q2) at (0, 2.25) {$Q_{3} \mapsfrom q_{2}$};
    \node[text=black] (q3) at (0, 1.5) {$Q_{5} \mapsfrom q_{3}$};
    \node[text=black] (q4) at (0, 0.75) {$Q_{7} \mapsfrom q_{4}$};
    \node[text=black] (q5) at (0, 0) {$Q_{0} \mapsfrom q_{5}$};
    
    \draw[thick, color=black] (q0.east) -- (4.75, 3.75) -- (5.75, 3.75); 
    \draw[thick, color=black] (q1.east) -- (4.75, 3) -- (5.75, 3);
    \draw[thick, color=black] (q2.east) -- (4.75, 2.25) -- (5.75, 2.25); 
    \draw[thick, color=black] (q3.east) -- (4.75, 1.5) -- (5.75, 1.5); 
    \draw[thick, color=black] (q4.east) -- (4.75, 0.75) -- (5.75, 0.75); 
    \draw[thick, color=black] (q5.east) -- (4.75, 0) -- (5.75, 0); 

    \draw[thick, color=black!10] (5.75, 3.75) -- (6.25, 3.75) node[text=black, midway, yshift=0mm] {$q_0$}; 
    \draw[thick, color=black!10] (5.75, 3) -- (6.25, 3) node[text=black, midway, yshift=0mm] {$q_1$}; 
    \draw[thick, color=black!10] (5.75, 2.25) -- (6.25, 2.25) node[text=black, midway, yshift=0mm] {$q_5$}; 
    \draw[thick, color=black!10] (5.75, 1.5) -- (6.25, 1.5) node[text=black, midway, yshift=0mm] {$q_3$}; 
    \draw[thick, color=black!10] (5.75, 0.75) -- (6.25, 0.75) node[text=black, midway, yshift=0mm] {$q_4$}; 
    \draw[thick, color=black!10] (5.75, 0) -- (6.25, 0) node[text=black, midway, yshift=0mm] {$q_2$}; 

    \draw[thick, color=black] (6.25, 3.75) -- (7.35, 3.75); 
    \draw[thick, color=black] (6.25, 3) -- (7.35, 3); 
    \draw[thick, color=black] (6.25, 2.25) -- (7.35, 2.25); 
    \draw[thick, color=black] (6.25, 1.5) -- (7.35, 1.5); 
    \draw[thick, color=black] (6.25, 0.75) -- (7.35, 0.75); 
    \draw[thick, color=black] (6.25, 0) -- (7.35, 0); 
    
    \filldraw [black] (1.25,3.75) circle (2pt);
    \draw[thick, color=black] (1.25,3.75) -- (1.25, 3);
    \node[inner sep=0pt] (cx3-targ) at (1.25, 3) {\normalsize$\bigoplus$};
    
    \filldraw [black] (2,3.75) circle (2pt);
    \draw[thick, color=black] (2,3.75) -- (2, 2.25);
    \node[inner sep=0pt] (cx3-targ) at (2, 2.25) {\normalsize$\bigoplus$};
    
    \filldraw [black] (2.75,3.75) circle (2pt);
    \draw[thick, color=black] (2.75,3.75) -- (2.75, 1.5);
    \node[inner sep=0pt] (cx3-targ) at (2.75, 1.5) {\normalsize$\bigoplus$};
    
    \filldraw [black] (3.5,3.75) circle (2pt);
    \draw[thick, color=black] (3.5,3.75) -- (3.5, 0.75);
    \node[inner sep=0pt] (cx3-targ) at (3.5, 0.75) {\normalsize$\bigoplus$};
    
    \draw[thin, rounded corners=10pt, dashed] (3.85, 4.1) rectangle (4.65, -0.35);
    \draw[IEEERed, fill=IEEERed!25, thick, rounded corners=8pt, fill opacity=.9] (3.95, 4) rectangle (4.55, -0.25);
    \filldraw [black] (4.25,3.75) circle (2pt);
    \draw[thick, color=black] (4.25,3.75) -- (4.25, 0);
    \node[inner sep=0pt] (cx3-targ) at (4.25, 0) {\normalsize$\bigoplus$};

    \draw[thin, rounded corners=10pt, dashed] (4.85, 4.1) rectangle (7.15, -0.35);
    \draw[IEEEGreen, fill=IEEEGreen!25, thick, rounded corners=8pt, fill opacity=0.9] (4.95, 2.5) rectangle (5.55, -0.25);
    \draw[thick] (5.25, 2.25) -- (5.25, 0);
    \node[] at (5.25, 2.25) {$\bigtimes$};
    \node[] at (5.25, 0) {$\bigtimes$};
    
    \draw[IEEEGreen, fill=IEEEGreen!25, thick, rounded corners=8pt, fill opacity=0.9] (6.45, 4) rectangle (7.05, 2);
    \filldraw [black] (6.75,3.75) circle (2pt);
    \draw[thick, color=black] (6.75,3.75) -- (6.75, 2.25);
    \node[inner sep=0pt] (cx3-targ) at (6.75, 2.25) {\normalsize$\bigoplus$};

    \draw[latex-, thick] (4.25, 4.2) to [bend left=30] (6, 4.2) node[anchor=west, yshift=2mm, xshift=-1mm] {replaces};
\end{tikzpicture}

%% file: figures/layers.tex
\begin{tikzpicture}[scale=1.0, font=\footnotesize]    
    \node[text=black] (q0) at (0, 3) {$q_0$};
    \node[text=black] (q1) at (0, 2) {$q_1$};
    \node[text=black] (q2) at (0, 1) {$q_2$};
    \node[text=black] (q3) at (0, 0) {$q_3$};

    \node[rectangle, draw=black, rounded corners=1pt, minimum size=0.8cm, thick, fill=white] (h00) at (1, 3) {$H$};
    \node[rectangle, draw=black, rounded corners=1pt, minimum size=0.8cm, thick, fill=white] (z10) at (1, 2) {$Z$};
    
    \node[rectangle, draw=black, rounded corners=1pt, minimum size=0.8cm, thick, fill=white] (h20) at (1, 1) {$H$};

    \node[circle, thick, fill=black, minimum size=4pt, inner sep=0pt] (cx0-ctrl) at (2, 3) {};
    \node[inner sep=0pt, fill=white] (cx0-targ) at (2, 2) {\normalsize$\bigoplus$};
    \draw[black, thick] (cx0-ctrl.south) -- (cx0-targ.north) node[midway, xshift=2mm] {\scriptsize g1};

    \node[circle, thick, fill=black, minimum size=4pt, inner sep=0pt] (cx1-ctrl) at (2, 1) {};
    \node[inner sep=0pt, fill=white] (cx1-targ) at (2, 0) {\normalsize$\bigoplus$};
    \draw[black, thick] (cx1-ctrl.south) -- (cx1-targ.north) node[midway, xshift=2mm] {\scriptsize g2};

    \node[circle, thick, fill=black, minimum size=4pt, inner sep=0pt] (cx2-ctrl) at (3, 2) {};
    \node[inner sep=0pt, fill=white] (cx2-targ) at (3, 1) {\normalsize$\bigoplus$};
    \draw[black, thick] (cx2-ctrl.south) -- (cx2-targ.north) node[midway, xshift=2mm] {\scriptsize g3};

    \node[rectangle, draw=black, rounded corners=1pt, minimum size=0.8cm, thick, fill=white] (h10) at (4, 2) {$H$};

    \node[circle, thick, fill=black, minimum size=4pt, inner sep=0pt] (cx3-ctrl) at (5.25, 3) {};
    \node[inner sep=0pt, fill=white] (cx3-targ) at (5.25, 2) {\normalsize$\bigoplus$};
    \draw[black, thick] (cx3-ctrl.south) -- (cx3-targ.north) node[midway, xshift=2mm] {\scriptsize g4};

    \draw[thick, color=black] (cx0-ctrl.west) -- (cx3-ctrl.east);

    \draw[thick, color=black, loosely dashed] (2.5, -0.45) -- (2.5, 3.45);
    \draw[thick, color=black, loosely dashed] (4.75, -0.45) -- (4.75, 3.45);

    \draw[thick, decorate,decoration={brace,amplitude=4pt, mirror}] (0.25, -0.5) -- (2.4, -0.5) node[midway, yshift=-5mm]{\scriptsize 1st Layer};
    \draw[thick, decorate,decoration={brace,amplitude=4pt, mirror}] (2.6, -0.5) -- (4.65, -0.5) node[midway, yshift=-5mm]{\scriptsize 2nd Layer};
    \draw[thick, decorate,decoration={brace,amplitude=4pt, mirror}] (4.8, -0.5) -- (5.95, -0.5) node[midway, yshift=-5mm]{\scriptsize 3rd Layer};

    \begin{pgfonlayer}{background}
    
    \draw[thick, color=black] (q0.east) -- (h00.west);
    \draw[thick, color=black] (q1.east) -- (z10.west);
    \draw[thick, color=black] (q2.east) -- (h20.west);
    \draw[thick, color=black] (q3.east) -- (cx1-targ.west);
    
    \draw[thick, color=black] (h00.east) -- (cx0-ctrl.west);
    \draw[thick, color=black] (z10.east) -- (cx0-targ.west);
    \draw[thick, color=black] (h20.east) -- (cx1-ctrl.west);

    \draw[thick, color=black] (cx0-targ.west) -- (cx2-ctrl.east);
    \draw[thick, color=black] (cx1-ctrl.west) -- (cx2-targ.east);
    \draw[thick, color=black] (cx1-targ.west) -- (6, 0);

    \draw[thick, color=black] (cx2-ctrl.east) -- (h10.west);
    \draw[thick, color=black] (cx2-targ.east) -- (6, 1);

    \draw[thick, color=black] (h10.east) -- (cx3-targ.west);
    \draw[thick, color=black] (cx3-ctrl.east) -- (6, 3);
    \draw[thick, color=black] (cx3-targ.east) -- (6, 2);
    
    \end{pgfonlayer}

\end{tikzpicture}

%% file: figures/expansion.tex
\begin{tikzpicture}[scale=1,font=\footnotesize]
\node[circle, draw=black, fill=IEEEGray!25, text=black, thick] (q0) at (-1, 1) {$Q_{0}$};
\node[circle, draw=black, fill=IEEEBlue!50, text=black, thick, minimum size=3mm] (q1) at (0, 1) {$Q_{1}$};
\node[circle, draw=black, fill=IEEEBlue!75, text=white, thick, inner sep=1pt] at (q1.north east) {\scriptsize$q_1$};

\node[circle, draw=black, fill=IEEEGray!25, text=black, thick, minimum size=3mm] (q2) at (1, 1) {$Q_{2}$};

\node[circle, draw=black, fill=IEEEGray!25, text=black, thick, minimum size=5mm] (q3) at (-1, 0) {$Q_{3}$};
\node[circle, draw=black, fill=IEEEGray!25, text=black, thick, minimum size=5mm] (q4) at (0, 0) {$Q_{4}$};
\node[circle, draw=black, fill=IEEEGray!25, text=black, thick, minimum size=5mm] (q5) at (1, 0) {$Q_{5}$};

\node[circle, draw=black, fill=IEEEBlue!50, text=black, thick, minimum size=5mm] (q6) at (-1, -1) {$Q_{6}$};
\node[circle, draw=black, fill=IEEEBlue!75, text=white, thick, inner sep=1pt] at (q6.north east) {\scriptsize$q_0$};

\node[circle, draw=black, fill=IEEEGray!25, text=black, thick, minimum size=5mm] (q7) at (0, -1) {$Q_{7}$};
\node[circle, draw=black, fill=IEEEGray!25, text=black, thick, minimum size=5mm] (q8) at (1, -1) {$Q_{8}$};

\draw[line width=1.5mm, color=IEEEDarkYellow] (q0.east) -- (q1.west);
\draw[very thick, color=IEEERed!75] (q0.east) -- (q1.west);
\draw[line width=1.5mm, color=IEEEDarkYellow] (q1.east) -- (q2.west);
\draw[very thick, color=IEEERed!75] (q1.east) -- (q2.west);

\draw[very thick, color=black] (q3.east) -- (q4.west);
\draw[very thick, color=black] (q4.east) -- (q5.west);

\draw[line width=1.5mm, color=IEEEDarkYellow] (q6.east) -- (q7.west);
\draw[very thick, color=IEEERed!75] (q6.east) -- (q7.west);
\draw[very thick, color=black] (q7.east) -- (q8.west);

\draw[very thick, color=black] (q0.south) -- (q3.north);
\draw[line width=1.5mm, color=IEEEDarkYellow] (q3.south) -- (q6.north);
\draw[very thick, color=IEEERed!75] (q3.south) -- (q6.north);

\draw[line width=1.5mm, color=IEEEDarkYellow] (q1.south) -- (q4.north);
\draw[very thick, color=IEEERed!75] (q1.south) -- (q4.north);
\draw[very thick, color=black] (q4.south) -- (q7.north);

\draw[very thick, color=black] (q2.south) -- (q5.north);
\draw[very thick, color=black] (q5.south) -- (q8.north);
\end{tikzpicture}

%% file: figures/astar.tex
\begin{tikzpicture}[scale=1, font=\footnotesize]

    \node[circle, fill=black, thick, text=white, draw=black, minimum size=8pt] (root) at (4, 0) {};

    \node[circle, fill=IEEEGray!25, thick, draw=black, inner sep=3pt] (0swap01) at (1, -3) {$Q_1, Q_0$};
    \node[circle, fill=IEEEGray!25, thick, draw=black, inner sep=3pt] (0swap12) at (2.5, -3) {$Q_1, Q_2$};
    \node[circle, fill=IEEEGray!25, thick, draw=black, inner sep=3pt] (0swap14) at (4, -3) {$Q_1, Q_4$};
    \node[circle, fill=IEEEGray!25, thick, draw=black, inner sep=3pt] (0swap63) at (5.5, -3) {$Q_6, Q_3$};
    \node[circle, fill=IEEEGray!25, thick, draw=black, inner sep=3pt] (0swap67) at (7, -3) {$Q_6, Q_7$};

    \node[inner sep=2pt, xshift=4mm, yshift=2mm] at (0swap01.north) {\scriptsize $f:2$};
    \node[inner sep=2pt, xshift=4mm, yshift=2mm] at (0swap12.north) {\scriptsize $f:4$};
    \node[inner sep=2pt, xshift=4mm, yshift=2mm] at (0swap14.north) {\scriptsize $f:2$};
    \node[inner sep=2pt, xshift=4mm, yshift=2mm] at (0swap63.north) {\scriptsize $f:2$};
    \node[inner sep=2pt, xshift=4mm, yshift=2mm] at (0swap67.north) {\scriptsize $f:2$};
    
    \draw[thick, black, -latex] (root) to[out=250, in=90] (0swap01);
    \draw[thick, black, -latex, dashed] (root) to[out=260, in=90] (0swap12);
    \draw[thick, black, -latex, dashed] (root) to[out=270, in=90] (0swap14);
    \draw[thick, black, -latex, dashed] (root) to[out=280, in=90] (0swap63);
    \draw[thick, black, -latex, dashed] (root) to[out=290, in=90] (0swap67);

    \node[circle, fill=IEEEGray!25, thick, draw=black, inner sep=3pt] (1swap03) at (1, -6) {$Q_0, Q_3$};
    \node[circle, fill=IEEEGray!25, thick, draw=black, inner sep=3pt] (1swap63) at (2.5, -6) {$Q_6, Q_3$};
    \node[circle, fill=IEEEGray!25, thick, draw=black, inner sep=3pt] (1swap67) at (4, -6) {$Q_6, Q_7$};

    \node[inner sep=2pt, xshift=4mm, yshift=2mm] at (1swap03.north) {\scriptsize $f:2$};
    \node[inner sep=2pt, xshift=4mm, yshift=2mm] at (1swap63.north) {\scriptsize $f:2$};
    \node[inner sep=2pt, xshift=4mm, yshift=2mm] at (1swap67.north) {\scriptsize $f:4$};

    \node[text=black, inner sep=1pt, anchor=center, yshift=2mm, xshift=-4mm] at (1swap03.north) {\faIcon{flag-checkered}};
    \node[text=black, inner sep=1pt, anchor=center, yshift=2mm, xshift=-4mm] at (1swap63.north) {\faIcon{flag-checkered}};
    
    \draw[thick, black, -latex, dashed] (0swap01) to[out=270, in=90] (1swap03);
    \draw[thick, black, -latex, dashed] (0swap01) to[out=280, in=90] (1swap63);
    \draw[thick, black, -latex, dashed] (0swap01) to[out=290, in=90] (1swap67);
\end{tikzpicture}

%% file: figures/wire-iterator.tex
\tikzset{
    wire1/.pic = {
        \node[circle, thick, draw=IEEECyan, fill=IEEECyan!40, minimum size=14pt, inner sep=0pt, label={\texttt{it}}] at (0, 1.5) {};
        \node[circle, draw=black, thick, minimum size=7pt, inner sep=0pt, fill=white ] (q0) at (0, 1.5) {};
        \node[circle, draw=black, thick, minimum size=7pt, inner sep=0pt, fill=white, opacity=.25] (q1) at (0, 0) {};

        \node[rectangle, minimum size=28pt, rounded corners=8pt, minimum height=68pt, minimum width=28pt, fill=IEEEGray!25, draw=black, thick] (cx) at (2, 0.75) {\rotatebox{90}{CTRL - X}};

        \node[circle, minimum size=28pt, fill=IEEEGray!25, draw=black, thick] (z) at (4.5, 1.5) {Z};

        \draw[-latex, thick] (q0.east) -- ([yshift=7.5mm]cx.west) node[midway,yshift=2mm] {\texttt{\%q0\_0}};
        \draw[-latex, thick] ([yshift=7.5mm]cx.east) -- (z.west) node[midway,yshift=2mm] {\texttt{\%q0\_1}};

        \draw[-latex, thick, color=black!50, dashed] (z.east) -- ++(0.75, 0);

        \draw[-latex, thick, color=black!25] (q1.east) --([yshift=-7.5mm]cx.west);
        \draw[-latex, thick, color=black!25] ([yshift=-7.5mm]cx.east) -- ++(3.25, 0);

        \draw[-latex, thick, dashed] ([xshift=-2mm]q0.south west) to[in=135, out=225] ++(0, -2.8) node[midway, xshift=-1.1cm] {\rotatebox{90}{\texttt{++it}}};
    }
}

\tikzset{
    wire2/.pic = {
        \node[circle, draw=black, thick, minimum size=7pt, inner sep=0pt, fill=white ] (q0) at (0, 1.5) {};
        \node[circle, draw=black, thick, minimum size=7pt, inner sep=0pt, fill=white, opacity=.25] (q1) at (0, 0) {};

        \node[rectangle, rounded corners=10pt, minimum height=74pt, minimum width=34pt, draw=IEEECyan, fill=IEEECyan!40, thick, label={\texttt{it}}] at (2, 0.75) {};
        \node[rectangle, rounded corners=8pt, minimum height=68pt, minimum width=28pt, fill=IEEEGray!25, draw=black, thick] (cx) at (2, 0.75) {\rotatebox{90}{CTRL - X}};

        \node[circle, minimum size=28pt, fill=IEEEGray!25, draw=black, thick] (z) at (4.5, 1.5) {Z};

        \draw[-latex, thick] (q0.east) -- ([yshift=7.5mm]cx.west) node[midway,yshift=2mm] {\texttt{\%q0\_0}};
        \draw[-latex, thick] ([yshift=7.5mm]cx.east) -- (z.west) node[midway,yshift=2mm] {\texttt{\%q0\_1}};

        \draw[-latex, thick, color=black!50, dashed] (z.east) -- ++(0.75, 0);

        \draw[-latex, thick, color=black!25] (q1.east) --([yshift=-7.5mm]cx.west);
        \draw[-latex, thick, color=black!25] ([yshift=-7.5mm]cx.east) -- ++(3.25, 0);

        \draw[-latex, thick, dashed] ([xshift=-2mm]q0.south west) to[in=135, out=225] ++(0, -2.8) node[midway, xshift=-1.1cm] {\rotatebox{90}{\texttt{++it}}};
    }
}

\tikzset{
    wire3/.pic = {
        \node[circle, draw=black, thick, minimum size=7pt, inner sep=0pt, fill=white ] (q0) at (0, 1.5) {};
        \node[circle, draw=black, thick, minimum size=7pt, inner sep=0pt, fill=white, opacity=.25] (q1) at (0, 0) {};

        \node[rectangle, rounded corners=8pt, minimum height=68pt, minimum width=28pt, fill=IEEEGray!25, draw=black, thick] (cx) at (2, 0.75) {\rotatebox{90}{CTRL - X}};

        \node[circle, thick, draw=IEEECyan, fill=IEEECyan!40, minimum size=34pt, inner sep=0pt, label={\texttt{it}}] at (4.5, 1.5) {};
        \node[circle, minimum size=28pt, fill=IEEEGray!25, draw=black, thick] (z) at (4.5, 1.5) {Z};

        \draw[-latex, thick] (q0.east) -- ([yshift=7.5mm]cx.west) node[midway,yshift=2mm] {\texttt{\%q0\_0}};
        \draw[-latex, thick] ([yshift=7.5mm]cx.east) -- (z.west) node[midway,yshift=2mm] {\texttt{\%q0\_1}};

        \draw[-latex, thick, color=black!25] (q1.east) --([yshift=-7.5mm]cx.west);
        \draw[-latex, thick, color=black!25] ([yshift=-7.5mm]cx.east) -- ++(3.25, 0);

        \draw[-latex, thick, color=black!50, dashed] (z.east) -- ++(0.75, 0);
    }
}

\begin{tikzpicture}[scale=1, font=\footnotesize]
\pic at (0, 6) {wire1};
\pic at (0, 3) {wire2};
\pic at (0, 0) {wire3};
\end{tikzpicture}

%% file: figures/layering.tex
\begin{tikzpicture}[scale=1.0, font=\footnotesize]

\node[circle, draw=black, thick, minimum size=7pt, inner sep=0pt, fill=white] (q0) at (-0.5, 3) {};
\node[circle, draw=black, thick, minimum size=7pt, inner sep=0pt, fill=white] (q1) at (-0.5, 1.5) {};
\node[circle, draw=black, thick, minimum size=7pt, inner sep=0pt, fill=white] (q2) at (-0.5, 0) {};
\node[circle, draw=black, thick, minimum size=7pt, inner sep=0pt, fill=white] (q3) at (-0.5, -1.5) {};
\node[circle, draw=black, thick, minimum size=7pt, inner sep=0pt, fill=white] (q4) at (-0.5, -3) {};

\node[circle, dashed, minimum size=28pt, fill=IEEEGray!25, draw=black, thick] (q0h) at (1.5, 3) {H};
\node[circle, dashed, minimum size=28pt, fill=IEEEGray!25, draw=black, thick] (q1h) at (1.5, 1.5) {H};

\node[circle, dashed, fill=IEEEGray!25, minimum size=28pt, draw=black, thick] (q0z) at (4, 3) {Z};

\draw[dashed, rounded corners=8pt] (-0.9, 2.4) rectangle (6.1, 3.6);
\draw[dashed, rounded corners=8pt] (-0.9, 0.9) rectangle (3.7, 2.1);
\draw[dashed, rounded corners=8pt] (-0.9, -0.6) rectangle (3.7, 0.6);
\draw[dashed, rounded corners=8pt] (-0.9, -2.1) rectangle (1.25, -0.9);
\draw[dashed, rounded corners=8pt] (-0.9, -2.4) rectangle (1.25, -3.6);

\node[rectangle, rounded corners=10pt, minimum height=74pt, minimum width=34pt, draw=IEEEGreen!75, fill=IEEEGreen!25, thick] (q3q4cx-highlight) at (1.5, -2.25) {};
\node[rectangle, rounded corners=8pt, minimum height=68pt, minimum width=28pt, fill=IEEEGray!25, draw=black, thick] (q3q4cx) at (1.5, -2.25) {\rotatebox{90}{CTRL - X}};

\node[rectangle, rounded corners=10pt, minimum height=74pt, minimum width=34pt, draw=IEEEGreen!75, fill=IEEEGreen!25, thick] (q1q2cx-highlight) at (4, 0.75) {};
\node[rectangle, rounded corners=8pt, minimum height=68pt, minimum width=28pt, fill=IEEEGray!25, draw=black, thick] (q1q2cx) at (4, 0.75) {\rotatebox{90}{CTRL - Y}};

\node[rectangle, rounded corners=8pt, minimum height=68pt, minimum width=28pt, fill=IEEEGray!25, draw=black, thick] (q0q1cx) at (6.5, 2.25) {\rotatebox{90}{CTRL - X}};

\draw[thick] (q0.east) -- (q0h.west) node[above, midway] {\texttt{\%q0\_0}};
\draw[thick] (q0h.east) -- (q0z.west) node[above, midway] {\texttt{\%q0\_1}};
\draw[-Bracket, thick] (q0z.east) -- ([yshift=7.5mm]q0q1cx.west) node[above, midway] {\texttt{\%q0\_2}};

\draw[thick] (q1.east) -- (q1h.west) node[above, midway] {\texttt{\%q1\_0}};
\draw[-Bracket, thick] (q1h.east) -- ([yshift=7.5mm]q1q2cx.west) node[above, midway] {\texttt{\%q1\_1}};

\draw[-Bracket, thick] (q2.east) -- ([yshift=-7.5mm]q1q2cx.west) node[above, midway] {\texttt{\%q2\_0}};;

\draw[-Bracket, thick] (q3.east) -- ([yshift=7.5mm]q3q4cx.west) node[above, midway] {\texttt{\%q3\_0}};

\draw[-Bracket, thick] (q4.east) -- ([yshift=-7.5mm]q3q4cx.west) node[above, midway] {\texttt{\%q4\_0}};;

\draw[-latex, black!50, thick, dashed] ([yshift=7.5mm]q3q4cx.east) -- ++(4, 0);
\draw[-latex, black!50, thick, dashed] ([yshift=-7.5mm]q3q4cx.east) -- ++(4, 0);

\draw[-Bracket, black!50, thick, dashed] ([yshift=7.5mm]q1q2cx.east) -- ([yshift=-7.5mm]q0q1cx.west);
\draw[-latex, thick, color=black!50, dashed] ([yshift=-7.5mm]q1q2cx.east) -- ++(1.5, 0);

\end{tikzpicture}

%% file: figures/two-qubit-block.tex
\begin{tikzpicture}[scale=1.0, font=\footnotesize]

\node[rectangle, rounded corners=8pt, minimum height=68pt, minimum width=28pt, fill=IEEEGray!25, draw=black, thick] (cx) at (0, 0.75) {\rotatebox{90}{CTRL - X}};

\node[circle, minimum size=28pt, fill=IEEEGray!25, draw=black, thick] (h0) at (1.5, 0) {H};
\node[circle, minimum size=28pt, fill=IEEEGray!25, draw=black, thick] (z0) at (3, 0) {Z};
\node[circle, minimum size=28pt, fill=IEEEGray!25, draw=black, thick] (h1) at (1.5, 1.5) {H};

\node[rectangle, rounded corners=8pt, minimum height=68pt, minimum width=28pt, fill=IEEEGray!25, draw=black, thick] (cz) at (4.5, 0.75) {\rotatebox{90}{CTRL - Z}};

\node[rectangle, rounded corners=8pt, minimum height=68pt, minimum width=28pt, fill=IEEEGray!25, draw=black, thick] (cy) at (6, -0.75) {\rotatebox{90}{CTRL - Y}};

\begin{pgfonlayer}{background}
    \node[draw=black, dashed, fill=IEEEGray!5, fit=(cx)(cz), rounded corners=8pt] (box) {};
  \end{pgfonlayer}

\draw[latex-, thick] ([yshift=7.5mm]cx.west) -- ++(-.5, 0);
\draw[latex-, thick] ([yshift=-7.5mm]cx.west) -- ++(-.5, 0);

\draw[-latex, thick] ([yshift=7.5mm]cx.east) -- (h1.west);
\draw[-latex, thick] ([yshift=-7.5mm]cx.east) -- (h0.west);

\draw[-latex, thick] (h1.east) -- ([yshift=7.5mm]cz.west);

\draw[-latex, thick] (h0.east) -- (z0.west);
\draw[-latex, thick] (z0.east) -- ([yshift=-7.5mm]cz.west);

\draw[-latex, thick, dashed] ([yshift=7.5mm]cz.east) -- ++(2, 0);
\draw[-latex, thick] ([yshift=-7.5mm]cz.east) -- ([yshift=7.5mm]cy.west);

\draw[-latex, thick] (-1, -1.5) -- ([yshift=-7.5mm]cy.west);
\draw[-latex, thick, dashed] ([yshift=7.5mm]cy.east) -- ++(.5, 0);
\draw[-latex, thick, dashed] ([yshift=-7.5mm]cy.east) -- ++(.5, 0);

\end{tikzpicture}

%% file: figures/reconstruction.tex
\begin{tikzpicture}[scale=1, font=\footnotesize]

\node[circle, fill=black, thick, text=white, draw=black, minimum size=8pt] (root) at (0, 0) {};

\node[isosceles triangle, draw=black, fill=IEEEGray!25, rotate=90, anchor=east, isosceles triangle apex angle=60, minimum size=18pt, opacity=.1, thick] (subtree-0) at (-1.5, -1.5) {};
\node[isosceles triangle, draw=black, fill=IEEEGray!25, rotate=90, anchor=east, isosceles triangle apex angle=60, minimum size=18pt, opacity=.1, thick] (subtree-1) at (3, -3.5) {};
\node[isosceles triangle, draw=black, fill=IEEEGray!25, rotate=90, anchor=east, isosceles triangle apex angle=60, minimum size=18pt, opacity=.1, thick] (subtree-2) at (1.5, -5.5) {};

\node[circle, fill=IEEEGray!25, thick, draw=black, inner sep=3pt] (swap01) at (1.5, -2) {$Q_0, Q_1$};
\node[circle, fill=IEEEGray!25, thick, draw=black, inner sep=3pt] (swap32) at (0, -4) {$Q_3, Q_2$};
\node[circle, fill=IEEEGray!25, thick, draw=black, inner sep=3pt] (swap43) at (-1.5, -6) {$Q_4, Q_3$};
\node[text=black, inner sep=1pt, anchor=center, yshift=2mm, xshift=-4mm] at (swap43.north) {\faIcon{flag-checkered}};

\draw[thick, black, -latex, opacity=.1] (root.south) to[out=270, in=90] (subtree-0.east);
\draw[thick, black, -latex, opacity=.1] (swap01.south) to[out=270, in=90] (subtree-1.east);
\draw[thick, black, -latex, opacity=.1] (swap32.south) to[out=270, in=90] (subtree-2.east);

\draw[thick, black, -latex] (root.south) to[out=270, in=90] (swap01);

\draw[thick, black, -latex] (swap01.south) to[out=270, in=90] (swap32.north);
\draw[thick, black, -latex] (swap32.south) to[out=270, in=90] (swap43.north);

\draw[very thick, IEEEBlue, -latex] (swap01.west) to[out=180, in=265] ([xshift=-1mm]root.south);
\draw[very thick, IEEEBlue, -latex] (swap32.east) to[out=0, in=275] ([xshift=1mm]swap01.south);
\draw[very thick, IEEEBlue, -latex] (swap43.east) to[out=0, in=275] ([xshift=1mm]swap32.south);

\end{tikzpicture}

%% file: figures/same-layout-tree.tex
\begin{tikzpicture}[scale=1, font=\footnotesize]
\node[circle, fill=black, thick, text=white, draw=black, minimum size=8pt] (root) at (0, 0) {};
\node[align=left, anchor=south west] at (root.north east) {
    p: (0 1 2)\\
    h: (0 1 2)
};

\node[circle, fill=IEEEGray!25, thick, draw=black, inner sep=3pt] (l-swap02) at (-1.5, -2) {$H_0, H_2$};
\node[align=left, anchor=east, inner sep=8pt] at (l-swap02.west) {
    p: (0 1 2)\\
    h: (2 1 0)
};

\node[isosceles triangle, draw=black, fill=IEEEGray!25, rotate=90, anchor=east, isosceles triangle apex angle=60, minimum size=18pt, opacity=.1, thick] (l-subtree-02) at (-2.5, -3.4) {};

\node[circle, fill=IEEEGray!25, thick, draw=black, inner sep=3pt] (l-swap01) at (-0.5, -4) {$H_0, H_1$};
\node[align=left, anchor=west, inner sep=8pt] at (l-swap01.east) {
    p: (0 1 2)\\
    h: (2 0 1)
};

\node[isosceles triangle, draw=black, fill=IEEEGray!25, rotate=90, anchor=east, isosceles triangle apex angle=60, minimum size=18pt, opacity=.1, thick] (l-subtree-01) at (-1.5, -5.4) {};

\node[circle, draw=IEEEDarkYellow, fill=IEEEDarkYellow!40, minimum size=38pt, inner sep=5pt] (l-swap12-highlight) at (0.5, -6) {};
\node[circle, fill=IEEEGray!25, thick, draw=black, inner sep=3pt] (l-swap12) at (0.5, -6) {$H_1, H_2$};
\node[align=left, anchor=west, inner sep=8pt] at (l-swap12-highlight.east) {
    p: (0 1 2)\\
    h: (1 0 2)
};

\node[circle, draw=IEEEDarkYellow, fill=IEEEDarkYellow!40, minimum size=38pt, inner sep=5pt] (r-swap01-highlight) at (1.5, -2) {};
\node[circle, fill=IEEEGray!25, thick, draw=black, inner sep=3pt] (r-swap01) at (1.5, -2) {$H_0, H_1$};
\node[align=left, anchor=west, inner sep=8pt] at (r-swap01-highlight.east) {
    p: (0 1 2)\\
    h: (1 0 2)
};

\draw[thick, black, -latex] (root.south) to[out=270, in=90] (l-swap02);
\draw[thick, black, -latex] (root.south) to[out=270, in=90] (r-swap01);

\draw[thick, black, -latex] (l-swap02.south) to[out=270, in=90] (l-swap01.north);
\draw[thick, black, -latex] (l-swap01.south) to[out=270, in=90] (l-swap12.north);

\draw[thick, black, -latex, opacity=.1] (l-swap02.south) to[out=270, in=90] (l-subtree-02.east);
\draw[thick, black, -latex, opacity=.1] (l-swap01.south) to[out=270, in=90] (l-subtree-01.east);
\end{tikzpicture}

    

%% file: figures/rewire.tex
\begin{tikzpicture}[scale=1.0, font=\footnotesize]

\node[circle, draw=black, thick, minimum size=7pt, inner sep=0pt, fill=white] (q0) at (0, 3) {};
\node[circle, draw=black, thick, minimum size=7pt, inner sep=0pt, fill=white] (q1) at (0, 1.5) {};
\node[circle, draw=black, thick, minimum size=7pt, inner sep=0pt, fill=white] (q2) at (0, 0) {};

\node[rectangle, rounded corners=8pt, minimum height=68pt, minimum width=28pt, fill=IEEEGray!25, draw=black, thick] (q1q2cx) at (2, 0.75) {\rotatebox{90}{CTRL - X}};

\draw[thick, draw=black, -latex] (q0.east) -- node[above] {\small\texttt{\%q0\_0}} ++({3.5cm+11pt}, 0);

\draw[thick, draw=black, -latex] (q1.east) -- node[above] {\small\texttt{\%q1\_0}} ([yshift=7.5mm]q1q2cx.west);
\draw[thick, draw=black, -latex] (q2.east) -- node[above] {\small\texttt{\%q2\_0}} ([yshift=-7.5mm]q1q2cx.west);

\draw[thick, draw=black, -latex] ([yshift=7.5mm]q1q2cx.east) -- node[above] {\small\texttt{\%q1\_1}} ++(1.5, 0);
\draw[thick, draw=black, -latex] ([yshift=-7.5mm]q1q2cx.east) -- node[above] {\small\texttt{\%q2\_1}} ++(1.5, 0);

\draw[thick, -latex, dashed] (5, 1.5) to[out=0, in=90] node[xshift=6mm, yshift=-1mm, midway, above] {SWAP} (6.25, 0);


\node[circle, draw=black, thick, minimum size=7pt, inner sep=0pt, fill=white] (q0) at (1.75, -1.5) {};
\node[circle, draw=black, thick, minimum size=7pt, inner sep=0pt, fill=white] (q1) at (1.75, -3) {};
\node[circle, draw=black, thick, minimum size=7pt, inner sep=0pt, fill=white] (q2) at (1.75, -4.5) {};

\node[rectangle, rounded corners=8pt, minimum height=28pt, minimum width=68pt, fill=IEEEGray!25, draw=black, thick, rotate=90] (swap) at (3.75, -2.25) {SWAP};

\draw[thick, draw=black, -latex] (q0.east) -- node[above] {\small\texttt{\%q0\_0}} ([yshift=7.5mm]swap.north);
\draw[thick, draw=black, -latex] (q1.east) -- node[above] {\small\texttt{\%q1\_0}} ([yshift=-7.5mm]swap.north);

\node[rectangle, rounded corners=8pt, minimum height=68pt, minimum width=28pt, fill=IEEEGray!25, draw=black, thick] (q1q2cx) at (6.25, -3.75) {\rotatebox{90}{CTRL - X}};

\draw[thick, draw=black, -latex] (q2.east) -- node[above] {\small\texttt{\%q2\_0}} ([yshift=-7.5mm]q1q2cx.west);
\draw[thick, draw=black, -latex] ([yshift=7.5mm]swap.south) to[out=0, in=180] node[above, xshift=6mm, yshift=-3mm] {\small\texttt{\%q0\_1}} ([yshift=7.5mm]q1q2cx.west);

\draw[thick, draw=black, -latex] ([yshift=-7.5mm]swap.south) to[out=0, in=180] ++(1.5, 1.5) -- node[above] {\small\texttt{\%q1\_1}} ++(2.5, 0);
\draw[thick, draw=black, -latex] ([yshift=7.5mm]q1q2cx.east) -- node[above] {\small\texttt{\%q1\_2}} ++(1.5, 0);
\draw[thick, draw=black, -latex] ([yshift=-7.5mm]q1q2cx.east) -- node[above] {\small\texttt{\%q2\_1}} ++(1.5, 0);

\end{tikzpicture}

%% file: plots/qmap-comparison-runtime.tex
\begin{tikzpicture}[font=\footnotesize]
\begin{loglogaxis}[
    enlargelimits=0.1,
    width=\textwidth, 
    height=\textwidth,
    xmin=1, xmax=100000,
    ymin=1, ymax=100000,
    ylabel={MQT Compiler Collection [ms]},
    xlabel={QMAP [ms]},
    axis equal image,
]
\addplot [
    thick, 
    black,
    dashed,
    domain=1:100000
] {x};

\addplot[
    only marks,
    color=IEEEBlue!75,
    fill opacity=0.25,
    mark=*,
    mark size=1.5pt,
    ]
table[col sep=comma, x={qmap_time}, y={mqt_time}]{data/qmap-comparison.csv};
\end{loglogaxis}
\end{tikzpicture}

%% file: plots/tket-comparison-runtime.tex
\begin{tikzpicture}[font=\footnotesize]
\begin{loglogaxis}[
    enlargelimits=0.1,
    width=\textwidth, 
    height=\textwidth,
    xmin=1, xmax=10000,
    ymin=1, ymax=10000,
    xlabel={TKET [ms]},
    axis equal image,
]
\addplot [
    thick, 
    black,
    dashed,
    domain=1:10000
] {x};

\addplot[
    only marks,
    color=IEEEBlue!75,
    fill opacity=0.25,
    mark=*,
    mark size=1.5pt,
    ]
table[col sep=comma, x={tket_time}, y={mqt_time}]{data/tket-comparison.csv};
\end{loglogaxis}
\end{tikzpicture}

%% file: plots/qiskit-comparison-runtime.tex
\begin{tikzpicture}[font=\footnotesize]
\begin{loglogaxis}[
    enlargelimits=0.1,
    width=\textwidth, 
    height=\textwidth,
    xmin=1, xmax=5000,
    ymin=1, ymax=5000,
    xlabel={Qiskit [ms]},
    axis equal image,
]
\addplot [
    thick, 
    black,
    dashed,
    domain=1:5000
] {x};

\addplot[
    only marks,
    color=IEEEDarkOrange!75,
    fill opacity=0.25,
    mark=*,
    mark size=1.5pt,
] table[col sep=comma, x={qiskit_time}, y={mqt_time}]{data/qiskit-fw-only-comparison.csv};

\addplot[
    only marks,
    color=IEEEBlue!75,
    fill opacity=0.25,
    mark=*,
    mark size=1.5pt,
] table[col sep=comma, x={qiskit_time}, y={mqt_time}]{data/qiskit-comparison.csv};
\end{loglogaxis}
\end{tikzpicture}

%% file: plots/qmap-comparison-nswaps.tex
\begin{tikzpicture}[font=\footnotesize]
\begin{loglogaxis}[
    enlargelimits=0.1,
    width=\textwidth, 
    height=\textwidth,
    xmin=20, xmax=10000,
    ymin=20, ymax=10000,
    ylabel={MQT Compiler Collection [\#SWAP]},
    xlabel={QMAP [\#SWAP]},
    axis equal image,
]
\addplot [
    thick, 
    black,
    dashed,
    domain=20:10000
] {x};

\addplot[
    only marks,
    color=IEEEBlue!75,
    fill opacity=0.25,
    mark=*, 
    mark size=1.5pt,
    ]
table[col sep=comma, x={qmap_nswaps}, y={mqt_nswaps}]{data/qmap-comparison.csv};
\end{loglogaxis}
\end{tikzpicture}

%% file: plots/tket-comparison-nswaps.tex
\begin{tikzpicture}[font=\footnotesize]
\begin{loglogaxis}[
    enlargelimits=0.1,
    width=\textwidth, 
    height=\textwidth,
    xmin=10, xmax=10000,
    ymin=10, ymax=10000,
    xlabel={TKET [\#SWAP]},
    axis equal image,
]
\addplot [
    thick, 
    black,
    dashed,
    domain=10:10000
] {x};

\addplot[
    only marks,
    color=IEEEBlue!75,
    fill opacity=0.25,
    mark=*, 
    mark size=1.5pt,
    ]
table[col sep=comma, x={tket_nswaps}, y={mqt_nswaps}]{data/tket-comparison.csv};
\end{loglogaxis}
\end{tikzpicture}

%% file: plots/qiskit-comparison-nswaps.tex
\begin{tikzpicture}[font=\footnotesize]
\begin{loglogaxis}[
    enlargelimits=0.1,
    width=\textwidth, 
    height=\textwidth,
    xmin=20, xmax=10000,
    ymin=20, ymax=10000,
    xlabel={Qiskit [\#SWAP]},
    legend pos=south east,
    legend style={font=\scriptsize},
    axis equal image,
]
\addplot [
    thick, 
    black,
    dashed,
    domain=20:10000,
    forget plot
] {x};

\addplot[
    only marks,
    color=IEEEDarkOrange!75,
    fill opacity=0.25,
    mark=*,
    mark size=1.5pt,
] table[col sep=comma, x={qiskit_nswaps}, y={mqt_nswaps}]{data/qiskit-fw-only-comparison.csv};
\addlegendentry{w/o refinement}

\addplot[
    only marks,
    color=IEEEBlue!75,
    fill opacity=0.25,
    mark=*, 
    mark size=1.5pt,
] table[col sep=comma, x={qiskit_nswaps}, y={mqt_nswaps}]{data/qiskit-comparison.csv};
\addlegendentry{w/ refinement}
\end{loglogaxis}
\end{tikzpicture}